\documentclass[tighten,twocolumn]{aastex631}

\usepackage{iftex}
\ifxetex
    
    \usepackage{fontspec}
    \usepackage{unicode-math,verbatim}
    
\else
    
    \usepackage[T1]{fontenc}
    
    \usepackage{newtxtext} 
    \usepackage{newtxmath} 
    \DeclareSymbolFont{cmletters}{OML}{cmm}{m}{it}
    \DeclareMathSymbol{v}{\mathalpha}{cmletters}{"76}
\fi

\usepackage{graphicx}

\usepackage{amsmath}
\usepackage{wrapfig}

\usepackage{tikz}

\usepackage{booktabs}

\ifxetex
\newcommand\bh{\mdsmblkcircle}
\renewcommand\bullet{\mdsmblkcircle}
\else
\newcommand\bh{\bullet}
\fi

\usepackage{float}

\begin{document}

\shortauthors{K{\i}ro\u{g}lu et al.}

\title{3D Hydrodynamic Simulations of Micro-Tidal Disruption Events Favor a Narrow, Ultra-relativistic Jet for GRB\,250702B}

\author[0000-0003-4412-2176]{Fulya K{\i}ro\u{g}lu}
\affil{Center for Interdisciplinary Exploration \& Research in Astrophysics (CIERA), Northwestern University, Evanston, IL 60201, USA}
\affil{NSF-Simons AI Institute for the Sky (SkAI), 172 E. Chestnut St., Chicago, IL 60611, USA}\email{fulya.kiroglu@northwestern.edu}

\author[0000-0003-2012-5217]{Taeho Ryu}
\affiliation{JILA, University of Colorado and National Institute of Standards and Technology, 440 UCB, Boulder, CO 80309-0440, USA}
\affiliation{Department of Astrophysical and Planetary Sciences, 391 UCB, University of Colorado, Boulder, CO 80309-0391, USA}

\author[0000-0002-9182-2047]{Alexander Tchekhovskoy}
\affil{Department of Physics \& Astronomy, Northwestern University, Evanston, IL 60208, USA}
\affil{Center for Interdisciplinary Exploration \& Research in Astrophysics (CIERA), Northwestern University, Evanston, IL 60201, USA}
\affil{NSF-Simons AI Institute for the Sky (SkAI), 172 E. Chestnut St., Chicago, IL 60611, USA}

\author[0000-0002-4086-3180]{Kyle Kremer}
\affiliation{Department of Astronomy \& Astrophysics, University of California, San Diego; La Jolla, CA 92093, USA}

\author[0000-0002-6347-3089]{Daichi Tsuna}
\affiliation{Center for Astrophysics $|$ Harvard \& Smithsonian, 60 Garden St, Cambridge, MA 02138, USA}

\author[0000-0002-4670-7509]{Brian D.~Metzger }
\affiliation{Department of Physics and Columbia Astrophysics Laboratory, Columbia University, New York, NY 10027, USA}
\affiliation{Center for Computational Astrophysics, Flatiron Institute, New York, NY 10010, USA}

\begin{abstract}

The longest detected $\gamma$-ray burst, GRB\,250702B, exhibited seven hours of prompt $\gamma$-ray emission, a day-long X-ray precursor, and a weeks-long fading X-ray tail. We show that a jetted micro-TDE can explain this unusual evolution by satisfying tight constraints imposed on the engine by the requirements of jet stability and observed energetics. Using 3D hydrodynamic \textsc{arepo} simulations of a $1M_\odot$ star disrupted by a $10M_\odot$ black hole, we find that within several hours the debris forms a quasi-steady envelope, with a low-density polar funnel ($\rho\propto r^{-2}$; half-opening angle $\approx15^\circ$). Applying an analytic jet stability model to the simulated envelope, we find that luminous ULGRBs operate their jets near the stability threshold. 
Thus, many ULGRB jets are too weak to break out and perish inside their envelopes, implying the existence of a class of low-luminosity ULGRBs powered by ``failed jets''. Powerful ($L_{\rm jet}\gtrsim{\rm few}\times10^{47}\,{\rm erg\,s^{-1}}$), ultra-relativistic ($\gamma_{\rm j}\sim100$), narrowly collimated ($\theta_{\rm b}\simeq 1^\circ$) jets can survive the current-driven kink instability, escape, and reproduce the observed high isotropic-equivalent luminosity of $\sim10^{51}\,{\rm erg\,s^{-1}}$. 
After breakout, jet widening ($\theta_{\rm b}\propto t$), combined with wind-suppressed accretion ($L_{\rm jet}\propto t^{-2}$), produces the steep early X-ray decline ($L_{\rm X,iso}\propto t^{-4}$); once the jet widens to the scale of the funnel walls, its opening angle saturates at $\simeq 10^\circ$, and the decline returns to the intrinsic $t^{-2}$. 
Together, our simulations and jet stability constraints favor narrow, highly relativistic jets as the engines of luminous ULGRBs and predict early jet breaks, motivating early multi-wavelength follow-up of future events.

\end{abstract}
 
\section{Introduction}
\label{sec:intro}

Gamma-ray bursts (GRB\,s) are among the most energetic transients in the universe, with isotropic-equivalent energies reaching $10^{55}\,
\mathrm{erg}$~\citep[for a review, see][]{Kumer2015}. Their extreme luminosities and tight collimation allow detection at cosmological distances, enabling the studies of high-redshift galaxies, reionization, heavy-element enrichment, and the first stars, while simultaneously offering direct insight into the most extreme astrophysical processes. 

The broad diversity of GRB\, timescales reflects the variety in both their progenitors and environments \citep[e.g.,][]{Bromberg2013}. 
However, only a small number of GRB\, progenitor channels has been robustly established. The collapse of massive stellar cores is widely accepted as the origin of long GRB\,s (lasting $\gtrsim 2$~s; e.g.,
\citealt{Hjorth2003,Stanek2003,Cano2017}), while mergers of compact objects power short GRB\,s (lasting $\lesssim 2$~s; e.g., \citealt{Berger2013,Tanvir2013,Abbott2017}), although the temporal distinction between these two sub-classes is now understood to be blurred \citep[e.g.,][]{Bromberg2013,Rastinejad2022}. Within this landscape, a small but growing class of events occupies the ``ultra-long'' regime, with prompt $\gamma$-ray emission persisting for up to $\sim 10^3\text{--}10^4$~s \citep[e.g.,][]{Gendre2013,Levan2014}. The progenitors of these ultra-long GRB\,s (ULGRB\,s) remain unknown. Beyond their exceptional durations, ULGRB\,s differ from the typical long GRB\,s in their temporal structure, quasi-regular recurrence, and locations within their host galaxies.

Several channels have been proposed to explain the long engine durations of ULGRB\,s, including the core collapse of low-metallicity blue supergiant stars \citep{Gendre2013,Perna2018,Tsuna2025}, the birth of a magnetar following massive-star collapse \citep{Greiner2015,Metzger+15}, the collapse of Population~III stars \citep{Nakauchi2012,Kinugawa2019}, tidal disruption of evolved helium stars by a stellar-mass BH or neutron star binary companion \citep{Fryer&Woosley98,Klencki&Metzger26,Villar+26}, tidal disruption of main-sequence stars by stellar-mass BHs, also known as \emph{micro-TDEs} (\citealt{Perets2016,Kremer2019_tde,Beniamini2025,OConnor2025}), and tidal disruption events by intermediate-mass BHs \citep{Krolik2011,Granot2025,Oganesyan2025,EylesFerris2026}.  
Multiwavelength observations are critical for distinguishing these channels. 
Although ULGRB\,s have relatively low peak fluxes compared to classical GRB\,s, their unusually long-lived and strongly variable X-ray and optical/radio light curves provide key diagnostics of the underlying central engine \citep[e.g.,][]{Stratta2013, Levan2014,Carney2025,OConnor2025,Goodwin2026}.

A striking recent example is GRB\,250702B \citep{Neights2026}, detected by \emph{Fermi} as the longest GRB\, on record, with a total duration exceeding $25,000$~s. The event exhibited episodic prompt emission consisting of three hard $\gamma$-ray episodes (cataloged as GRB\,s~250702B, D, and~E) spread over approximately $3.2$~hours, with quasi-regular spacing of $\sim 2,825$~s \citep{Levan2025}, and was preceded by a soft X-ray flare detected nearly a day earlier by \emph{Einstein Probe} \citep{Cheng2025}. When combined with the day-long X-ray precursor, the extraordinary and episodic prompt duration, and the apparent absence of a supernova counterpart, and large host-galaxy offset (5.7 kpc) \citep{Carney2025}, the phenomenology of GRB\,250702B is in tension with core-collapse-related scenarios, motivating consideration of compact-object-based progenitors distributed more broadly across host galaxies.

Among a broader class of progenitor models, micro-TDEs  are strong candidates for producing the extreme ULGRB\, behavior, as they naturally give rise to long-lived fallback accretion at relatively low luminosities and generate extended debris envelopes that can delay or episodically modulate jet emergence \citep[e.g.,][]{Perets2016,Kremer2022,Ryu2022,Kiroglu2023,An2025,Beniamini2025}. In addition, micro-TDEs are expected to occur at roughly comparable rates across a range of dynamically evolved environments, including globular clusters \citep{Perets2016, Kremer2019_tde}, nuclear star clusters \citep{Fragione2021,Rose2022}, young massive clusters \citep{Kremer2021_fbot,kiroglu2025a,Kiroglu2025c,Rastello2026}, and stellar triples undergoing Kozai--Lidov oscillations \citep{Fragione2019, Naoz2025}. 
The combination of long accretion timescales, and their occurrence in environments distributed broadly across host galaxies, makes micro-TDEs a compelling candidate progenitor for ULGRB\,s like GRB\,250702B.

While micro-TDEs are well-motivated ULGRB\, progenitors on the basis of their engines and host environments, whether such systems observable jetted transients has not yet been addressed.  Previous studies of micro-TDEs have established that a generic outcome is a BH embedded within a thick, rapidly rotating gas torus, qualitatively similar to a collapsar
\citep[e.g.,][]{Kremer2022,Ryu2022,Vynatheya2024}. 
If collapsars can power long GRB\,s, can micro-TDEs, with lower accretion rates, give rise to similar but dimmer and longer-lived GRB\,s? If the disrupting BH spins rapidly and the accretion flow formed in a micro-TDE can efficiently amplify and/or accumulate the large-scale magnetic flux near the BH, the hole can launch relativistic jets~\citep{Blandford&Znajek77} and power prompt $\gamma$-ray and/or X-ray emission. 
Although isolated stellar collapse generally forms slowly rotating BHs~\citep[e.g.,][]{FullerMa2019}, they can acquire non-negligible spin through a variety of channels. In isolated binaries, tidal torques and mass transfer can spin up the progenitor or the newly formed BH \citep[e.g.,][]{Qin_2018, MaFuller2019, Bavera_2020, FullerLu2022}. In dense stellar environments, accretion ~\citep[e.g.,][]{Kremer2018_xrb,Lopez2019, Kiroglu2025b,Kiroglu2025c,Newton2026,Roupas2026,Rozner2026} and stellar mergers~\citep[e.g.,][]{Kremer2020,Tsuna2025, Satish2026} can provide the spin-up. However, these channels result in spin distributions that depend sensitively on the uncertain stellar-evolution processes and accretion efficiencies, and remain poorly constrained. In contrast, hierarchical black-hole mergers produce spinning BHs with a well-characterized distribution peaking near $a_\bh \sim 0.7$~\citep[e.g.,][]{Berti_2008,Tichy_2008,Kesden_2010,AntoniniRasio2016,Fishbach_2017,Rodriguez2016,Mai2026}. Additionally, active galactic nuclei (AGN) disks provide an additional channel in which a combination of hierarchical mergers and gas-driven accretion produces spinning BHs~\citep[e.g.,][]{Vajpeyi2022,Yang2022, McKernan2024}.

Beyond the BH spin that determines whether a jet can be launched, the jet's ability to survive its passage through the surrounding debris is crucial for the observability of ULGRB\,s. Whether the micro-TDE debris settles into a geometry that provides a low-density escape path for the jet has not been quantitatively established, and is the central question addressed in this work. In this Letter, we study jet formation and escape in micro-TDEs as a way to produce ULGRB\,s.  We perform 3D hydrodynamic simulations of a $1\,M_\odot$ star disrupted by a $10\,M_\odot$ BH, extract the density profiles of the surrounding envelope, and apply an analytic jet stability framework based on these profiles. 

The paper is organized as follows. Section~\ref{sec:methods} describes the hydrodynamic simulations of micro-TDEs. Section~\ref{sec:tde} presents the micro-TDE disk evolution model based on viscous accretion and wind-driven mass loss. Section~\ref{sec:jet} presents the observational signatures of jetted micro-TDEs and assesses jet stability
during propagation through the debris envelope. Section~\ref{sec:250702B} applies our micro-TDE accretion model to GRB\,250702B, constraining the BH spin and beaming angle required to reproduce the observed prompt $\gamma$-ray luminosity and the X-ray light curve. We summarize our results and discuss their implications in Section~\ref{sec:conclusions}.

\section{Hydrodynamics Modeling}
\label{sec:methods}

We perform 3D hydrodynamic simulations using the moving-mesh code \textsc{arepo} \citep{Arepo,ArepoHydro,Weinberger2020}.
\textsc{arepo} inherits the advantages of both Eulerian finite-volume and Lagrangian smoothed particle methods, including shock capturing without artificial viscosity, minimal advection errors, efficient handling of supersonic flows, and an adaptive adjustment of spatial resolution. We adopt the Helmholtz equation of state \citep{Timmes2000}, which includes radiation pressure under the assumption of local thermodynamic equilibrium with composition inherited from the MESA progenitor model described in Section~\ref{sec:stellar_model} and advected with the flow without nuclear burning.

We model the BH as a point particle of mass $M_\bh = 10\,M_\odot$ that interacts with the gas only gravitationally and neither accretes any mass nor produces any radiative feedback. We then model in post-processing the subsequent (unresolved) disk evolution and compute the time-dependent accretion rate onto the BH (Section~\ref{sec:tde}), using the fallback rate $\dot{M}_{\rm fb}(t)$ computed from the \textsc{arepo} simulations following
\citet{Kremer2022,Kremer2023}. The fallback rate is constructed from the return-time distribution of the bound debris, where the return time of each cell is its Keplerian orbital period computed from the specific energy relative to the BH. Because a cell's specific energy is still evolving in the early post-pericenter phase, we assign the Keplerian period only after half an orbit, once the specific energy has stabilized.

In these simulations, it is crucial to resolve the disk structure as close to the BH as possible. We therefore employ adaptive mesh refinement within a shell of radius $2\times 10^4\,r_{\rm g}$ around the BH, where $r_{\rm g} = GM_\bh/c^2$ is the BH gravitational radius. We suppress the refinement within an inner region, $r < 100r_{\rm g}$, to prevent excessive cell splitting there, which would otherwise drive the timestep to prohibitively small values; in practice, this region defines the innermost resolved radius of our accretion flow. Outside this inner region and within the refinement shell, we split cells whose density exceeds $2\times 10^{-9}$\,g\,cm$^{-3}$, mass exceeds $6\times 10^{18}$\,g, or size-to-radius ratio $\Delta d/r > 0.2$; the density criterion restricts refinement to physically meaningful gas rather than vacuum-filled regions, and the mass floor prevents a runaway creation of low-mass cells.  The size-to-radius criterion ensures a minimum spatial resolution of $\Delta d \lesssim 0.2\,r$ throughout the refinement shell and $N\sim 10^{3-4} $ cells lie within $10^4\,r_{\rm g}$ of the BH.  At each time step we also de-refine cells with mass below $1.5\times 10^{18}$\,g and $\Delta d/r < 0.05$, so that the cell mass within the refinement shell never drops below this floor. The refinement scheme increases the total cell count from $\simeq 2.5\times 10^5$ at the start of the simulation to $\simeq 5\times 10^6$ within a few days, concentrating resolution within $\sim 10^4\,r_{\rm g}$ of the BH.

\begin{figure*}
    \centering
\includegraphics[width=0.8\linewidth]{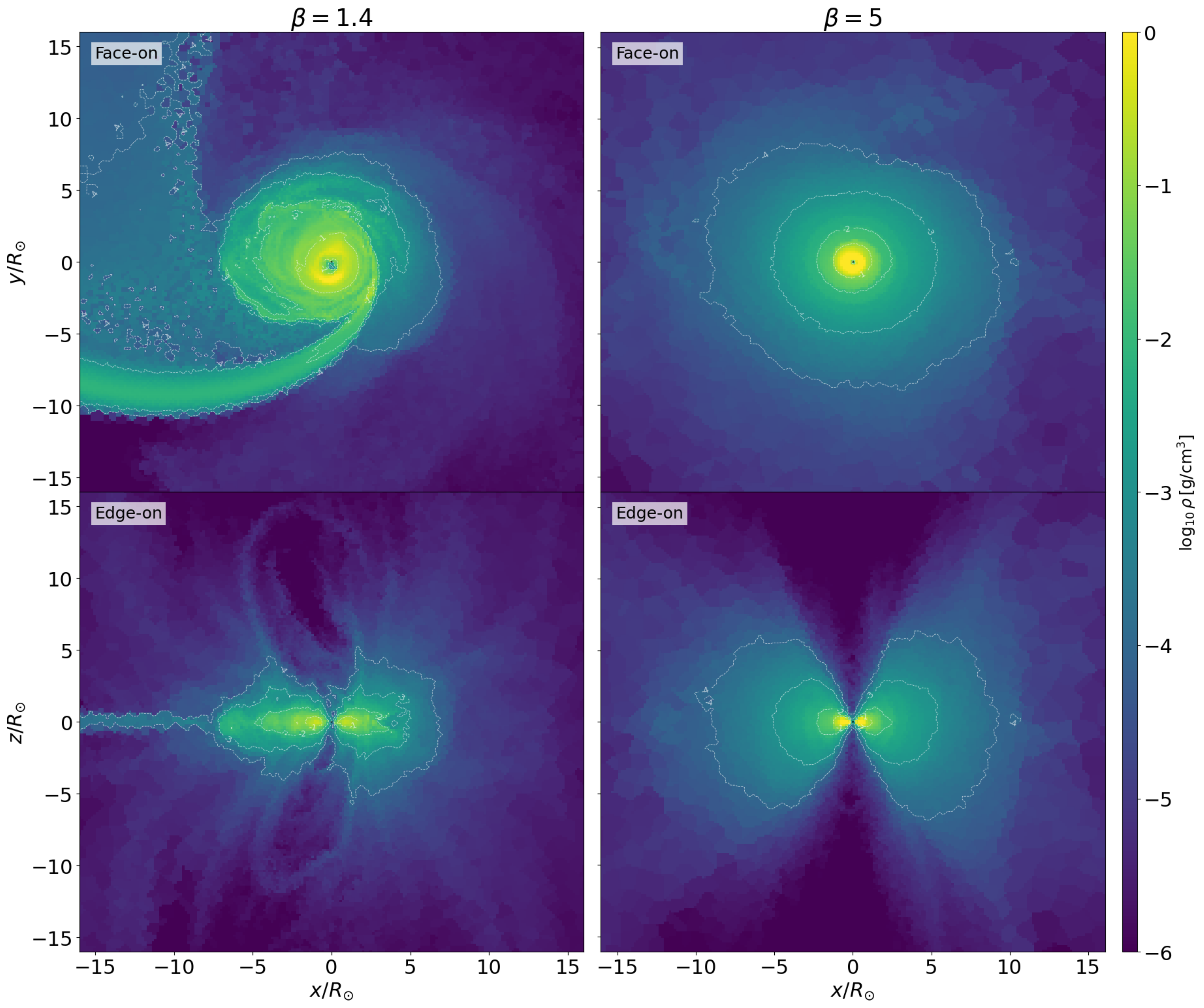}
        \caption{
        Horizontal (top) and vertical (bottom) density slices through \textsc{arepo} simulations at $t = 0.5$\,day of a $10\,M_\odot$ BH tidally disrupting a $1\,M_\odot$ star reveal the formation of an envelope with low-density polar funnels, which are conducive to relativistic jet formation and escape, for two different encounter strengths. 
        \textbf{[left column]:} Grazing encounter ($\beta \equiv r_{\rm T}/r_{\rm p} = 1.4$) forms an extended disk of radius $r_{\rm d} \approx 3\,R_\odot$.  Ongoing debris fallback onto the disk in the grazing encounter leads to polar funnel asymmetries and can result in jet wobbling and quasi-periodic variability.
        \textbf{[right column]:} Deep encounter ($\beta = 5$) forms a more compact and denser disk of radius $r_{\rm d} \approx 1\,R_\odot$. White contours mark consecutive orders of magnitude evenly spaced in $\log_{10}\rho$. Both encounters develop bipolar low-density polar funnels, with comparable half opening angles, $\theta_{\rm f}\approx 15^\circ$, within which $r^2\rho$ is approximately constant along the polar direction.}
        
\label{fig:snapshots}
\end{figure*}
\subsection{Stellar Model and Orbit}
\label{sec:stellar_model}
We create a $1\,M_\odot$ MESA model \citep{Paxton2015} with hydrogen core mass fraction of $0.5$ (corresponding to an age of $2.5$\,Gyr) and metallicity of $Z = 1\,Z_\odot$. The 1D MESA profile is mapped onto a 3D \textsc{arepo} grid with initial $N \simeq 2.5\times 10^5$ cells following the procedure of \citet{Ohlmann2017}: particles are distributed on nested Hierarchical Equal Area isoLatitude Pixelization (HEALPix) shells \citep{Gorski2005}, yielding an equal-mass, isotropic sampling with smaller cells in the dense core and larger cells in the envelope.

We interpolate the density and pressure from the MESA profile onto each cell and obtain the specific internal energy from the equation of state. Outside the stellar surface, we assign negligible ``vacuum'' values to the density and pressure. We adopt a cubic domain of side length $\approx 500\,R_\odot$, about a few hundred times the tidal disruption radius, and apply periodic boundary conditions. The domain is large enough that the bulk of the stellar debris remains well within the boundary over the duration of our simulations ($\lesssim 1$\,day), so the boundary condition does not affect the disk formation and evolution.   We relax the mapped star into hydrostatic equilibrium before initiating the disruption. 

We consider two initial orbits, both leading to the full disruption of the star: a grazing encounter with the penetration factor $\beta=1.4$ and semi-major axis  $a = 3\,R_{\odot}$, and a deep encounter with
$\beta=5$ and $a = 1\,R_{\odot}$, where $\beta \equiv r_{\rm T}/r_{\rm p} > 1$, $r_{\rm T}$ is the tidal disruption radius and $r_{\rm p} \equiv a(1-e)$ is the pericenter distance. In both cases the orbital eccentricity is $e = 0.5$ and the star begins at apocenter $r_{\rm apo} \equiv a(1+e)$. For bound orbits, whether the star is fully disrupted or undergoes stable eccentric mass transfer is set primarily by the pericenter distance \citep[e.g.,][]{Chen2026}. We therefore focus on varying the pericenter distance, which sets the depth of the tidal disruption and controls the compression of the star, and the corresponding disk properties such as size and thickness \citep[e.g.,][]{GuillochonRamirezRuiz2013,Stone2013,Ryu2020}. Eccentricity primarily affects the orbital binding energy of the debris, changing the fraction that remains bound to the BH and the efficiency of circularization into a disk \citep[e.g.,][]{Hayasaki2013, Bonnerot2016,Park2020}. In the case of nearly circular orbits, the star instead undergoes gradual ``tidal peeling'', a slow, repeated stripping of the stellar envelope over many orbits \citep{Xin2023}.

The bound orbit with our adopted moderate eccentricity ($e=0.5$) results in
about $80\%$ of the stellar mass being captured onto the BH, which favors the disk mass and jet energetics relevant to a luminous event like GRB\,250702B. 
 Such bound, moderately eccentric orbits could arise through several dynamical pathways: 3D hydrodynamic simulations show that binary--single encounters in dense star clusters can leave one star bound to the BH, in some cases on a close, low-eccentricity orbit that leads to a micro-TDE \citep{Ryu+2023,Ryu+2023b,Ryu+2024b}; and in AGN disks, dynamical friction from the disk gas can harden a star--BH binary until the star is disrupted \citep[e.g.,][]{Yang2022,Xin2023}. Alternatively, in the binary stellar evolution channel, an existing BH--star binary can be driven to disruption when a supernova natal kick brings the pericenter within the tidal radius \citep[e.g.,][]{Beniamini2025}.

The presence or absence of a supernova counterpart could help distinguish among the possible formation channels. In the dynamical formation channel, the BH forms well before it acquires and disrupts the companion, so any supernova long predates the micro-TDE and leaves no signature at the time of the event. A near-coincident supernova would be expected only in the binary-evolution scenario, where a natal kick directly places the star within the tidal radius; this configuration is also the least likely to produce the deeply-plunging encounter we adopt. The supernova non-detection in GRB250702 \citep{Gompertz2026} therefore supports a dynamical micro-TDE origin. However, it is possible that a supernova associated with the binary-evolution channel was present but remained undetected, given the source's large distance and the significant dust extinction in its host \citep{Carney2025}, so this channel cannot be firmly excluded.
\begin{figure}
    \centering    \includegraphics[width=1.\linewidth]{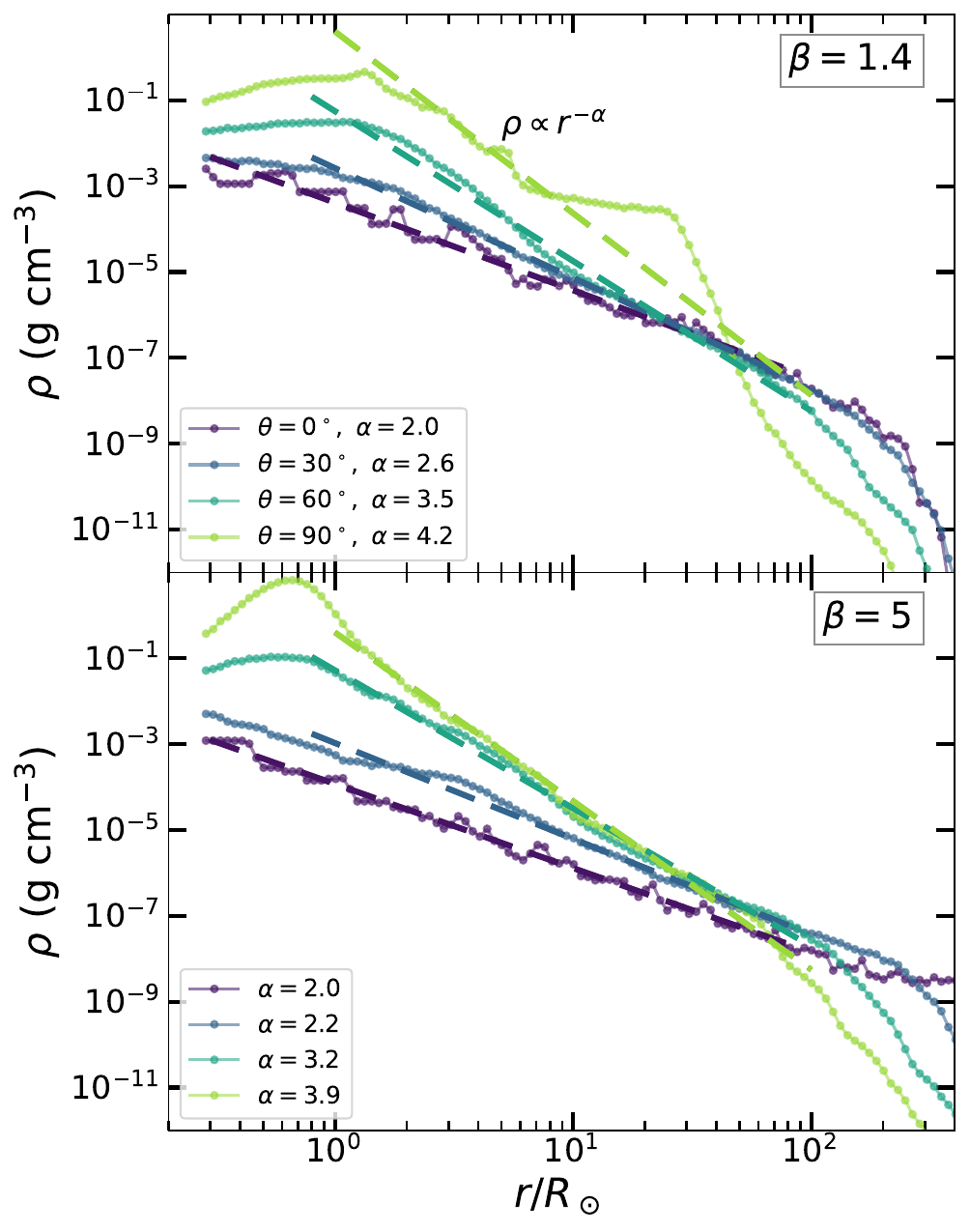}
    \caption{Radial density profiles of the debris envelope extracted from the \textsc{arepo} simulations at four polar angles at $t=4$~hours post-disruption, where $\theta = 0^\circ$ corresponds to the polar axis (low-density funnel) and $\theta = 90^\circ$ to the equatorial plane (disk midplane). Each profile is obtained by sampling the density at fixed polar angle $\theta$ as a function of radius from the BH, averaged over azimuthal angles and both hemispheres to suppress interpolation noise, with the density at each sample point taken from the nearest Voronoi cell. The top panel shows the grazing encounter ($\beta = 1.4$) and the bottom panel the deep encounter ($\beta = 5$); color encodes polar angle, and dashed lines indicate power-law fits, $\rho \propto r^{-\alpha}$, with best-fit slopes given in the legends. For the stronger encounter, the envelope is denser by a factor of several across all angles. While the polar profile follows $\rho \propto r^{-2}$, the profiles steepen toward lower latitudes ($\alpha > 2$) as the debris concentrates near the midplane.}
    \label{fig:density_profiles}
\end{figure}

\subsection{Disk formation and properties}
\label{sec:disk}

Our \textsc{arepo} simulations follow the disruption dynamics and the initial formation of the accretion disk within a day, corresponding to $5$--$10$ orbital times at the circularization radius, $r_{\rm c}$ \citep[e.g.,][]{Bonnerot2016}. Within this window, the inner disk ($r \lesssim r_{\rm c}$) circularizes and settles into a rotationally supported configuration, while the outer disk remains in the process of assembly from returning fallback debris. We do not attempt to capture the subsequent viscous evolution of the disk directly within \textsc{arepo}: at these radii the \textsc{arepo} scheme is not intended to model $\alpha$-viscosity transport, and the timestep near the BH makes extending the simulation to many viscous timescales computationally prohibitive. Instead, we extract the fallback rate $\dot{M}_{\rm fb}(t)$ from the \textsc{arepo} simulations.
This fallback rate then serves as input to the semi-analytic disk evolution model (Section~\ref{sec:tde}), where wind-modified viscous accretion drives disk drainage and radial spreading over the $\sim$~day to $\sim$~week timescales relevant to the observations.

Figure~\ref{fig:snapshots} shows face-on (top) and edge-on (bottom) density slices of the debris envelope formed after the disruption of a $1M_\odot$ star by a $10M_\odot$ BH at $t =0.5$\,day, for two penetration factors: a grazing encounter ($\beta=1.4$; left panels) and a deep encounter ($\beta=5$; right panels). The deep encounter produces a compact, dense inner disk ($r_{\rm d} \approx 1R_\odot$), while the grazing encounter forms a more extended configuration ($r_{\rm d} \approx 3R_\odot$). Both encounters develop a well-defined bipolar funnel with half-opening angle $\theta_{\rm f}\approx 15^\circ$ that can serve as a convenient escape funnel for BH powered relativistic jets.

To quantify the angular structure of this funnel, we show the radial density profiles extracted from our \textsc{arepo} simulations at four polar angles in Figure~\ref{fig:density_profiles}, where $\theta = 0^\circ$ corresponds to the low-density polar funnel along the spin axis and $\theta = 90^\circ$ to the dense equatorial plane of the debris disk. We compare grazing  and strong encounters in each panel. The density contrast between the polar funnel and the equatorial plane spans more than three orders of magnitude at $r \sim R_\odot$, from $\rho \sim 10^{-4}$\,g\,cm$^{-3}$ at $\theta = 0^\circ$ to $\rho \gtrsim 10^{-1}$\,g\,cm$^{-3}$ at $\theta = 90^\circ$. The polar density profile is well described by a power-law $\rho \propto r^{-2}$, while at intermediate and equatorial angles the profiles steepen towards $\rho \propto r^{-3}$. The polar $r^{-2}$ scaling is the key input to our jet propagation and stability calculations: it provides a steadily declining confining pressure that collimates rather than disrupts the jet \citep[e.g.,][]{Bromberg2016,Tchekhovskoy2016} which we discuss in Section~\ref{sec:jet_stability}.

\section{Micro-TDE Disk Evolution}
\label{sec:tde}
Our \textsc{arepo} simulations presented in Section~\ref{sec:methods} follow the stellar disruption and disk formation phases but do not capture mass accretion onto the BH. In this section, we develop an analytic model for the subsequent viscous accretion by incorporating fallback supply, viscous drainage onto the BH, and wind-driven mass loss. We later use the time-dependent accretion rate onto the BH to compute the jet power and the resulting X-ray light curve (Section~\ref{sec:jet}).

We consider a general model for the tidal disruption and accretion of a star of mass $M_* = m_{*,1}M_\odot$ and radius $R_* = R_\odot m_{*,1}$ onto a BH of mass $M_\bh = 10m_{\bh,10}M_\odot$ and dimensionless spin $a_\bh < 1$, which in general can be misaligned with the angular momentum axis of the accreted material. Depending on the values of $M_*$, $M_\bh$, and $a_\bh$, this model can apply to a variety of scenarios other than micro-TDEs, such as post--common-envelope binary mergers.

The disrupted star forms the fallback debris and accretion disk of the characteristic size set by the circularization radius, 
\begin{equation}
r_{\rm c} \simeq 2r_{\rm p} =  \frac{2 r_{\rm T}}{\beta}
\approx 2 \times 10^{5} r_{\rm g}\,\beta^{-1}\,
m_{*,1}^{2/3}\,m_{\bh,10}^{-2/3},
\end{equation}
where $r_{\rm p}$ is the pericenter radius, $r_{\rm T} \equiv R_{\star}\,(M_\bh/M_{\star})^{1/3}$ is the tidal radius, $\beta  \equiv r_{\rm T}/r_{\rm p} > 1$ is the penetration factor,  and $r_{\rm g} \equiv GM_\bh/c^2$ is the BH gravitational radius.

Adopting a standard $\alpha$-prescription \citep[][]{ShakuraSunyaev1973} we calculate the viscous accretion timescale of a geometrically thick disk with aspect ratio $h \equiv 0.5 h_{0.5}\equiv H/R$ for the fiducial parameter choices as
\begin{equation}
t_{\rm v} \simeq \alpha_{\rm v}^{-1}\,h^{-2}
\left( \frac{r_{\rm c}}{r_{\rm g}} \right)^{3/2} \frac{r_{\rm g}}{c}
\approx 2\,{\rm day}\, \beta^{-3/2}\,\alpha_{\rm v,-1}^{-1}\,h_{0.5}^{-2}\, m_{*,1},
\label{eq:tv}
\end{equation}
where $\alpha_{\rm v} = 0.1\,\alpha_{\rm v,-1}$
is the viscosity parameter.

Since the viscous timescale characterizes the global disk evolution, we evaluate it at $r \sim r_{\rm c}$ where the disk is well-defined and rotationally supported. From our \textsc{arepo} simulations (Figure~\ref{fig:disk_profiles} in Appendix), we see that the two encounters produce different aspect ratios near the circularization radius: the deep encounter ($\beta = 5$) yields $h \approx 0.5$, characteristic of geometrically thick, advection-dominated accretion flows at highly super-Eddington rates \citep[e.g.,][]{Narayan1994, Abramowicz2013, Blaes2014}, which are expected to eject most of the disk mass as winds \citep[e.g.,][]{Narayan1994, Blandford1999, StrubbeQuataert2009, 2025PhRvD.112l3044L}. The grazing encounter ($\beta = 1.4$) instead settles into a thinner, more rotationally supported configuration with $h \approx 0.2$. Because the viscous timescale scales as $t_{\rm v} \propto h^{-2}$, it is  approximately six times longer than in the grazing case than in the deep case, at the same radius. This slower viscous evolution shifts the accretion peak to later times and lower amplitudes, delaying the onset of the asymptotic $\dot{M}_{\rm acc}\propto t^{-2}$ decline. In what follows, we find that the $\beta = 5$ configuration provides a better match to the late-time X-ray decline of GRB\,250702B (Section~\ref{sec:250702B}).

\label{sec:disk_model}

Following \citet{Metzger2008,Kremer2019_tde,Kremer2023}, we approximate the disk mass distribution as a single ring at radius $r_{\rm d}$ where the surface density peaks, and track its mass and angular momentum evolution under fallback supply $\dot{M}_{\rm fb}$, viscous accretion onto the BH, and wind-driven mass loss. The disk mass evolves as
\begin{equation}
    \label{eq:Mdotd}
    \dot{M}_{\rm d} = -f\,\frac{M_{\rm d}}{t_{\rm v}} + \dot{M}_{\rm fb}.
\end{equation}
We adopt the order-unity factor, $f = 1$ (see \citealt{Metzger2008} for discussion of potentially more precise values for these parameters which lead to factor of order unity corrections).

Following \citet{Blandford1999},  we include mass loss from disk winds 
by adopting a radius-dependent inflow rate $\dot{M}_{\rm in}(r) = (r/r_{\rm d})^s\, M_{\rm d}/t_{\rm v}$, where $0 \le s \le 1$ parametrizes the radial slope of the wind mass outflow. We assume that the wind starts at the radius $r_{\rm acc}=10r_{\rm g} $ so that the inflow rate at this radius,

\begin{equation}
    \label{eq:Mdotacc}
    \dot{M}_{\rm acc} = \left(\frac{r_{\rm acc}}{r_{\rm d}}\right)^{\!s}\,
    \frac{ M_{\rm d}}{t_{\rm v}},
\end{equation}
gives the BH accretion rate.
The wind carries away the remainder, $\dot{M}_{\rm out} = [1 - (r_{\rm acc}/r_{\rm d})^s]\,  M_{\rm d}/t_{\rm v}$. The disk radius evolves as \citep[][]{Kremer2023}
\begin{equation}
    \label{eq:Rdot}
    \dot{r}_{\rm d} = \frac{2 r_{\rm d}}{t_{\rm v}}\Bigg[
    1 - C\!\left(1 - \left(\frac{r_{\rm acc}}{r_{\rm d}}\right)^{\!s}\right)
    + \left(\sqrt{\frac{r_{\rm c}}{r_{\rm d}}} - 1\right)
    \frac{\dot{M}_{\rm fb}\,t_{\rm v}}{M_{\rm d}}\Bigg],
\end{equation}
where $C = 2s/(2s+1)$ is the torque coefficient assuming the wind carries away angular momentum with its local specific value but exerts no net torque \citep{StonePringle2001, Kumar2008, 2024ApJ...965..175M}. We solve Equations~(\ref{eq:Mdotd}) and (\ref{eq:Rdot}) numerically to compute $M_{\rm d}(t)$ and $r_{\rm d}(t)$, where we take $\dot{M}_{\rm fb}(t)$ directly from our \textsc{arepo} simulations (Sec.~\ref{sec:disk}). As the disk radius $r_{\rm d}$ increases, the viscous accretion timescale increases accordingly $t_{\rm v} \propto r_{\rm d}^{3/2}$. We account for this evolving timescale when solving for the coupled evolution of the disk mass and radius. Inserting these time-dependent quantities into Equation~(\ref{eq:Mdotacc}) then yields the time-dependent accretion rate onto the BH.

At late times, the system asymptotes to the following scalings~\citep{Metzger2008, Kremer2019_tde,Kremer2023}, 
\begin{align}
    r_{\rm d} &\propto t^{2/3}, 
    \label{eq:scaling1}\\
    \dot{M}_{\rm d} &\propto t^{-(2s+4)/3} = t^{-5/3},
    \label{eq:scaling2}\\
    \dot{M}_{\rm acc} &\propto t^{-4(s+1)/3} = t^{-2},
    \label{eq:scaling3}
\end{align}
where the decay of $\dot{M}_{\rm acc}$, steeper than the canonical $t^{-5/3}$ TDE fallback rate, is the direct consequence of viscous disk spreading combined with wind-driven mass loss. We identify this $\dot{M}_{\rm acc}\propto t^{-2}$ scaling in Section~\ref{sec:250702B} as the origin of the late-time X-ray decay of GRB\,250702B. The $s=0.5$ value is supported by recent simulations of radiatively inefficient accretion, including neutron star post-merger accretion disks~\citep{2019MNRAS.482.3373F,2019MNRAS.490.4811C}, advection-dominated flows \citep{2012MNRAS.423.3083M,2012MNRAS.426.3241N,2013MNRAS.436.3856S,Cho2025} and Bondi accretion with weak cooling \citep{Guo2025}.

Recent general relativistic magnetohydrodynamic (GRMHD) simulations, which maintain continuous bidirectional causal connection between the BH and feeding scales and feature sufficiently large scale separation (e.g., reach inflow equilibrium over three orders of magnitude in distance), have brought nuance into this picture. They have revealed that $s$ is not a universal constant of accretion but depends on the state of the accretion flow~\citep{2025PhRvD.112l3044L}. When large-scale vertical magnetic flux fills the BH to capacity and becomes as strong as the gravitational force acting on the inner disk, it erupts from the BH, rips through the disk, and buoyantly rises away from the hole~\citep{Narayan+03,Tchekhovskoy2011}. This magnetically arrested disk (MAD) state features ordered accretion, strong magnetically-powered jets, and $s \approx 0.66\pm0.03$~\citep{2025PhRvD.112l3044L}. MAD-powered outflows can scramble the angular momentum of the system and lead to a rocking accretion disk (RAD) state. The RAD state features chaotic accretion, continuously reorienting weak jets, and $s\approx 0.87\pm0.05$~\citep{2025PhRvD.112l3044L}.

\section{JETTED MICRO-TDE OBSERVATIONAL SIGNATURES}
\label{sec:jet}

\subsection{Jet power and efficiency}
Large-scale magnetic field threading a rotating BH can extract its spin energy and launch twin Poynting-flux-dominated jets of electromagnetic luminosity,
\begin{equation}
    \label{eq:L_d}
    L_{\rm jet} = \eta_{\rm jet} (a_\bh) \dot{M}_{\rm acc} c^2,
\end{equation}
via the \citet[][BZ hereafter]{Blandford&Znajek77} process. Here, $\dot{M}_{\rm acc}$ is the BH accretion rate, for which we adopt the expression given by Equation~(\ref{eq:Mdotacc}), and $a_\bh$ is the dimensionless BH spin. The jet energy efficiency, defined by eq.~\eqref{eq:L_d} as the ratio of the extracted electromagnetic energy to accreted rest-mass energy, has been calibrated by GRMHD simulations~\citep[e.g.,][]{Tchekhovskoy+10,Tchekhovskoy2011,2012JPhCS.372a2040T,2015ASSL..414...45T},
\begin{equation}
\eta_{\rm jet}(a_\bh) \simeq 2.64\left(\frac{\varphi_\bh}{50}\right)^{2}\omega_{\rm H}^{2}f(\omega_{\rm H}),
\label{eq:etaBZ}
\end{equation}
where $\omega_{\rm H}= a_\bh/\left(1+(1-a_\bh^{2})^{1/2}\right)$ is the dimensionless angular frequency of the BH and $f(\omega_{\rm H}) \simeq 1+0.35 \omega_{\rm H}^{2}-0.58 \omega_{\rm H}^{4}$ is a high-spin correction. Here, $\varphi_\bh \equiv \Phi_\bh/(\dot M r_{\rm g}^2 c)^{1/2}$ is the dimensionless BH magnetic flux, which saturates at $\varphi_\bh\sim 50$ in the MAD state, and $\Phi_\bh=0.5\oiint_{r=r_{\rm H}} |B^r| \, \mathrm{d}A$ is the absolute BH magnetic flux, with the integral taken over both hemispheres of the event horizon, $r_{\rm H}=r_{\rm g}\left(1+(1-a_\bh^2)^{1/2}\right)$, and the factor of $0.5$ converting it to one hemisphere~\citep{Tchekhovskoy2011}. 

The black-hole spin is held fixed throughout our calculation as spin evolution is negligible over the engine lifetime. Integrating the accretion rate, the BH gains $\Delta M \sim \dot{M}_{\rm acc}\times 1\,{\rm day} \sim 0.02\,M_\odot$, corresponding to $\Delta M/M_\bullet \sim 2\times10^{-3}$. A significant spin change in a MAD flow requires accreting of order $10\%$ of the black-hole mass; adopting a spin-up parameter $(da/d\ln M_\bullet) \sim -10$ for a thick MAD near $a_\bullet \sim 1$ \citep{Lowell2024}, the corresponding change is $\Delta a \sim -2\times10^{-2}$, well below the adopted value. Spin extraction by the jet is bounded by the same mass budget and is similarly negligible.

Twin relativistic jets focus their radiation into a narrow solid angle $\Omega_{\rm b} = 4\pi(1-\cos\theta_{\rm b})$ around the disk rotational axis. Here, $\theta_{\rm b}$ is the beaming angle, which is set by the larger of the jet opening angle, $\theta_{\rm j}$ and the relativistic beaming angle, $1/\gamma_{\rm j}$: $\theta_{\rm b} = \max(\theta_{\rm j}, 1/\gamma_{\rm j})$. Which term dominates switches at breakout: while the jet is confined, it is strongly collimated, so $1/\gamma_{\rm j} > \theta_{\rm j}$ and $\theta_{\rm b} = 1/\gamma_{\rm j}$, whereas after breakout the jet widens and $\theta_{\rm j} > 1/\gamma_{\rm j}$, so $\theta_{\rm b} = \theta_{\rm j}$. An observer aligned with the jet axis therefore infers an isotropic-equivalent luminosity that is enhanced relative to the true twin jet power by the inverse of the beaming fraction, $f_{\rm b} \equiv \Omega_{\rm b}/4\pi = 1-\cos\theta_{\rm b}$. The isotropic-equivalent luminosity is therefore
\begin{equation}
    L_{\rm iso}(t) = \eta_{\rm rad} \, \eta_{\rm jet} f_{\rm b}^{-1}\, \dot{M}_{\rm acc} (t) c^2,
    \label{eq:Liso}
\end{equation}
where $\eta_{\rm rad}$ is the fraction of the jet power that is radiated in X-ray or $\gamma$-ray band, which converts the jet power into the observed radiative luminosity.

Why do we need to make the distinction between $\theta_{\rm j}$ and $\theta_{\rm b}$? If we used $\theta_{\rm j}$ for computing the beaming fraction, we would expect that a narrowly collimated jet, $\theta_{\rm j}=0.01\,\text{rad}\approx 6^\circ$, always boosts its luminosity by a factor, $1/f_{\rm b} = 1/(1-\cos\theta_{\rm j})\approx 2/\theta_{\rm j}^2 = 2\times 10^4$, for an on-axis observer. However, this is true only for fast jets, with $\gamma_{\rm j} \theta_{\rm j} \gtrsim 1$. If the jet moves at a lower Lorentz factor, e.g., $\gamma_{\rm j} = 10$, then its relativistic beaming angle much exceeds the opening angle, $\theta_{\rm b} = 1/\gamma_{\rm j} = 0.1 \gg \theta_{\rm j}$, and the true luminosity boost comes out to a much lower value, $200$, or a factor of $(\gamma_{\rm j}\theta_{\rm j})^{-2} = 10^2$ smaller.

\subsection{Jet Direction and Ambient Medium}
\label{sec:jet_direction}

In general, the BH spin axis can be misaligned with the angular momentum axis of the debris disk, depending on the initial spin--orbit misalignment of the BH--star orbit. In such a case, the disk undergoes solid-body Lense--Thirring precession with period \citep[e.g.,][]{Fragile2007,Stone2012},
\begin{equation}
P_{\rm LT} \approx
\frac{\pi}{5 a_\bh} \frac{r_{\rm g}}{c}
\left( \frac{r_{\rm d}}{r_{\rm g}} \right)^{5/2}
\left( \frac{r_{\rm ISCO}}{r_{\rm g}} \right)^{1/2},
\end{equation}
where $r_{\rm ISCO} \approx 3.4 r_{\rm g}$ corresponds to $a_\bh = 0.7$ for prograde orbit. Numerically for $s=0.5$ and $r_{\rm d} \propto t^{2/3}$,
\begin{equation}
P_{\rm LT} \approx
1.8 \times 10^{4}\,{\rm day}\,
\beta^{-5/2} a_{\bh,0.7}^{-1}
m_{*,1}^{5/3} m_{\bh,10}^{-2/3}
\left( \frac{t}{t_{\rm v}} \right)^{5/3}.
\end{equation}
Lense--Thirring torques are therefore not strong enough to cause the precession of the jets over the jet engine lifetime ($\sim$ day). Physically, this is because micro-TDE disks feature significant radial extents ($r_{\rm d}\sim 10^5 r_{\rm g}$) that imply large angular momentum content and long precession periods.  Recent GRMHD simulations show that magnetic torques from powerful jets can, in some regimes, reorient the inner disk angular momentum into alignment with the BH spin direction on timescales comparable to or shorter than the accretion timescale \citep[e.g.,][]{Mckinney2013,Chatterjee2025}. However, the alignment is limited to the immediate vicinity of the BH, $r\lesssim100r_{\rm g}$, which is orders of magnitude smaller than the simulated length scales. As a result, tilted BH systems with extended accretion disks in micro-TDEs launch jets along the rotational axis of the disk irrespective of the initial BH spin--disk misalignment. We therefore assume the configuration in which the jets propagate through the low-density polar funnel along the rotational axis of the extended disk.

\subsection{Jet Stability}
\label{sec:jet_stability}

The propagation of a Poynting-dominated relativistic jet through a confining medium is subject to current-driven (kink) instabilities that can disrupt the jet and cause it to stall before breakout. We can quantify the stability of the jet through the stability parameter~\citep{Bromberg2016, Tchekhovskoy2016},
\begin{equation}
    \Lambda
    = 2\frac{\gamma_{\rm j}\theta_{\rm j}}{0.03}
    = K \left(\frac{L_{\rm 1,jet}}{\rho_{\rm a}\,z_{\rm h}^2\,\gamma_{\rm j}^2\,c^3}
        \right)^{\!1/6},
    \label{eq:Lambda}
\end{equation}
where $\gamma_{\rm j}$ is the bulk Lorentz factor, where $L_{\rm 1,jet} = 0.5 L_{\rm jet}$ is one-sided jet power, $z_{\rm h}$ is the distance to the jet head, $\rho_{\rm a}$ is the ambient density, and $K = 20\,(2\pi/9)^{1/2}\,[\pi(5-\alpha)(3-\alpha)/6]^{1/3}$ is a constant numerical prefactor that depends on the radial slope of the ambient density, $\alpha \equiv -\mathrm{d}\log \rho_{\rm a}/\mathrm{d}\log r$. Our simulations have $\alpha = 2$ along the polar direction, as seen in Figure~\ref{fig:density_profiles}. 

We assume the jet is highly magnetized, with magnetization, $\sigma \equiv b^2/(4\pi\rho c^2) \gg 1$, and moves at $\beta_{\rm j} \equiv v_{\rm j}/c \simeq 1$, corresponding to Lorentz factor $\gamma_{\rm j}  \gg 1$; where $b$ is the comoving magnetic field strength. 
These conditions are expected for a Poynting-dominated jet launched by the Blandford--Znajek mechanism from a rapidly spinning BH in the MAD state: magnetic flux threading the BH horizon powers the jet, keeping it magnetically dominated at least near the jet launching point ($\sigma \gg 1$), and the resulting Alfv\'en speed (which sets the kink-mode growth rate) approaches $c$.

\begin{figure}
    \centering
     \includegraphics[width=1\linewidth]{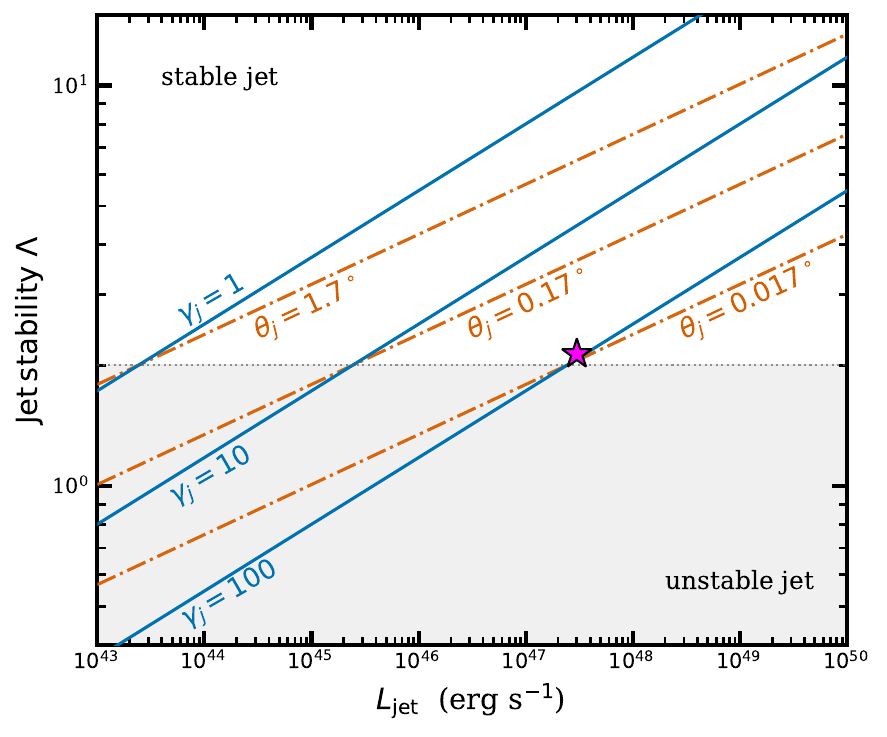}
\caption{
  Jet stability parameter, $\Lambda$, as a function of intrinsic jet power, $L_{\rm jet}$, for the deep encounter, $\beta=5$; $\Lambda$ is independent of distance for our radial density profile, $\rho\propto r^{-2}$. Blue solid curves show $\Lambda$ parameterized by Lorentz factor, $\gamma_{\rm j}=1,\,10,\,100$; orange dash-dotted curves show the same relation at fixed jet half-opening angle, $\theta_{\rm j}=1.7^\circ,\,0.17^\circ,\,0.017^\circ$, related by the condition, $\Lambda\equiv2\gamma_{\rm j}\theta_{\rm j}/0.03=2$, at their crossing points, so that each $\gamma_{\rm j}$ curve and its paired $\theta_{\rm j}$ curve cross at $\Lambda=2$, which marks the stability threshold. Jets above it ($\Lambda>2$) propagate stably, while those in the gray shaded region ($\Lambda<2$) are disrupted by the kink instability. The magenta star marks the deep encounter model with $L_{\rm jet}\simeq3\times 10^{47}\,{\rm erg\,s^{-1}}$ for $a_\bh=0.9$ and $\theta_{\rm j}=0.02^\circ$ ($\theta_{\rm b}=0.7^\circ$), which lies right above the jet stability boundary.}
    \label{fig:jet_stability}
\end{figure}

Because the polar density follows $\rho_{\rm a} \propto r^{-2}$, the factor of $\rho_{\rm a} z_{\rm h}^2$ in Equation~\eqref{eq:Lambda} is independent of the jet-head distance, and so is $\Lambda$; the stability result therefore holds at all radii and does not rely on the jet reaching a particular breakout height. Our profile sits precisely at the critical slope $\alpha = 2$ separating the two regimes identified by \citet{Bromberg2016}: in steeper profiles ($\alpha > 2$) the jet gradually opens up and stabilizes as it propagates, whereas in flatter profiles ($\alpha < 2$) it gradually collimates and grows increasingly unstable. Figure~\ref{fig:jet_stability} shows the jet stability parameter, $\Lambda$, evaluated at a fixed jet head distance, $z_{\rm h} = 10\,R_\odot$, as a function of the intrinsic jet luminosity, $L_{\rm jet}$, for three representative Lorentz factor values, $\gamma_{\rm j} = 1, 10, 100$, and for three representative jet opening angles, $\theta_{\rm j} = 0.017^\circ$, $0.17^\circ$, and $1.7^\circ$. (Note that we chose this particular value of $z_{\rm h}$ without loss of generality, because the density profile is such that $\rho_{\rm a}z_{\rm h}^2=\text{constant}$ and hence $\Lambda$ is independent of $z_{\rm h}$.) We compute the polar ambient density $\rho_{\rm a}$ directly from our \textsc{arepo} simulations. 

Reading the stability threshold ($\Lambda>2$) off Figure~\ref{fig:jet_stability},
the most relativistic jet ($\gamma_{\rm j}=100$), which is most susceptible
to the kink instability, requires the largest jet power to remain stable,
$L_{\rm jet,min}\simeq3\times10^{47}$~erg~s$^{-1}$. Producing this
power from the micro-TDE wind-modified ($s=0.5$) accretion rate
$\dot{M}_{\rm acc}\sim10^{-7}\,M_\odot\,{\rm s}^{-1}$ requires a rapidly spinning
BH, $a_\bh\gtrsim0.9$, and hence a jet efficiency
$\eta_{\rm jet}\gtrsim1$. Because slower, less relativistic jets have
progressively lower stability thresholds (the $\gamma_{\rm j}=1$ and $10$ loci
cross $\Lambda=2$ at correspondingly lower $L_{\rm jet}$, and the orange
dash-dotted contours give the associated opening angle
$\theta_{\rm j}=0.015\,\Lambda/\gamma_{\rm j}$), a jet launched by a BH
with the same jet power lies in the stable region ($\Lambda>2$) across the full
range $\gamma_{\rm j}\lesssim100$ considered here.  
Jet survival across the range of Lorentz factors and jet geometries considered in this work therefore makes micro-TDEs a viable central engine for ULGRB\,s.

\section{Application to GRB\,250702B}
\label{sec:250702B}

GRB\,250702B/EP\,250702a exhibits three temporally and spectrally distinct phases. \textit{Einstein Probe} detected a soft X-ray precursor approximately one day before the main burst
\citep{Cheng2025}, followed by three hard $\gamma$-ray episodes registered by \textit{Fermi}-GBM with quasi-regular spacing of $\sim 2,825$\,s \citep{Levan2025, Neights2026}. The \textit{Einstein Probe}  localization enabled continued monitoring with \textit{Swift}, NuSTAR, and \textit{Chandra} \citep{OConnor2025, EylesFerris2026}, revealing a long-lived X-ray counterpart fading  over days to weeks. The late time X-ray light curve of EP\,250702a exhibits two observationally distinct phases. The first is the EP-WXT phase, lasting several hours and contemporaneous with the Fermi $\gamma$-ray triggers, during which the X-ray emission peaks at $L_{\rm X,iso} \sim 10^{49}~\mathrm{erg~s^{-1}}$ with a hard spectrum ($\Gamma \simeq 0.2$, \citealt{Cheng2025}). The second is the EP-FXT/\emph{Swift}/\emph{Chandra} phase, beginning after the final \emph{Fermi} trigger and decaying as $L_{\rm X} \propto t^{-1.9}$ with a softer spectrum ($\Gamma \simeq 1.8$, \citealt{OConnor2025}).

\begin{figure*}
    \centering
\includegraphics[width=0.8\linewidth]{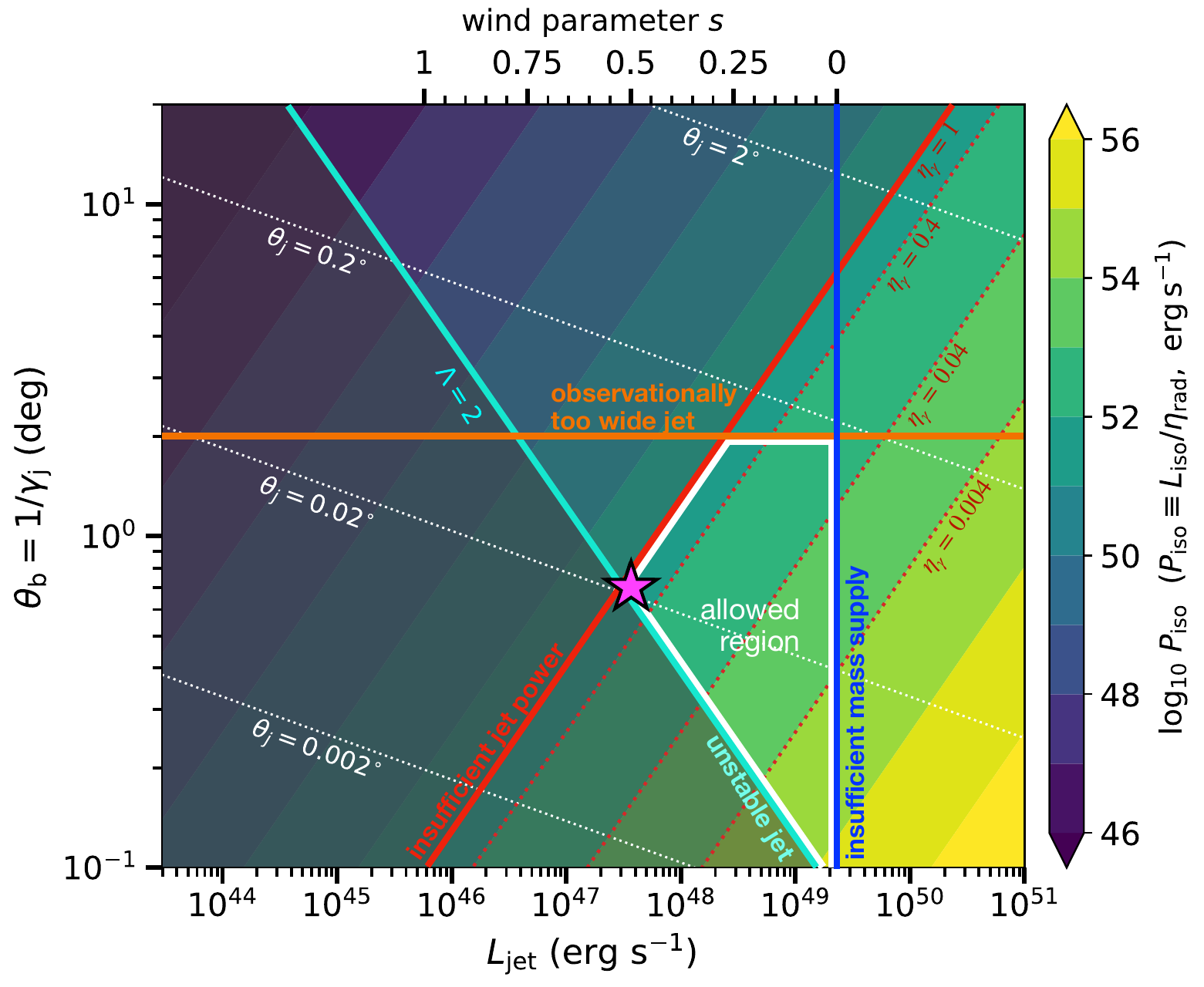}
\caption{
    Jet power and jet stability requirements constrain our micro-TDE model of GRB\,250702 to lie to the right of the solid red and cyan lines, respectively, as seen on the color map of the logarithm of isotropic-equivalent jet power, $P_{\rm iso}$,
    as a function of the intrinsic jet power,
    $L_{\rm jet}$, and beaming angle, $\theta_{\rm b}$ (see color bar). The upper $x$-axis shows the values of the wind mass loss parameter, $s$, that result in the intrinsic jet power values, for black hole spin of $a_\bh =0.9$. The magenta star marks the minimum power solution,
    $L_{\rm jet}\simeq 3\times10^{47}$~erg/s and $\theta_{\rm b}=0.7^\circ$ that is consistent with all available constraints.
    The solid red line, $P_{\rm iso}=10^{51.6}\,{\rm erg\,s^{-1}}$, matches GRB\,250702 isotropic equivalent $\gamma$-ray luminosity \citep{Neights2026} for unity radiative efficiency ($\eta_\gamma=1$; dotted red lines do so for lower $\eta_\gamma$ values, as labeled) and rules out the shaded \emph{insufficient jet power} region.   
    The solid cyan curve, $\Lambda=2$, rules out the shaded \emph{unstable jet} region ($\Lambda<2$), where the 3D magnetic kink instability disrupts the jets. The horizontal orange line reflects the observational upper limit on the jet opening angle, $\max\theta_{\rm b}\simeq2^\circ$, which ensures that the jet break happens early enough to be missed by the radio observations \citep{Goodwin2026}. The vertical blue line marks the maximum jet power allowed by the mass supply (i.e., assuming optimistically that the accretion disk does not have any wind mass loss, $s = 0$). Dotted white lines indicate jet opening angle values before the breakout (as labeled). Together, these constraints leave the allowed parameter space enclosed by the white solid lines.}
     
       \label{fig:Piso}
\end{figure*}

We propose that the $\gamma$-ray and soft X-ray emission from GRB\,250702B originates from a jetted micro-TDE in which a rapidly spinning stellar-mass BH disrupts a companion main-sequence star. This interpretation is motivated by three observational features that are difficult to reconcile with a standard collapsar or blue-supergiant scenario: the day-long soft X-ray precursor with no natural analog in collapsar models, the absence of an accompanying supernova signature, and the multi-hour duration of the prompt emission. The $5.7$\,kpc offset from the host galaxy nucleus \citep{Carney2025} disfavors a jetted tidal disruption around a supermassive BH, leaving open a range of stellar-mass and intermediate-mass BHs. However, the rapid sub-second variability of the prompt $\gamma$-ray emission points to a stellar-mass BH origin as the minimum variability timescale sets an upper limit on the central engine size $\lesssim 100\,M_{\odot}$ \citep{Neights2026}. 

\begin{figure*}
    \centering
\includegraphics[width=0.8\linewidth]{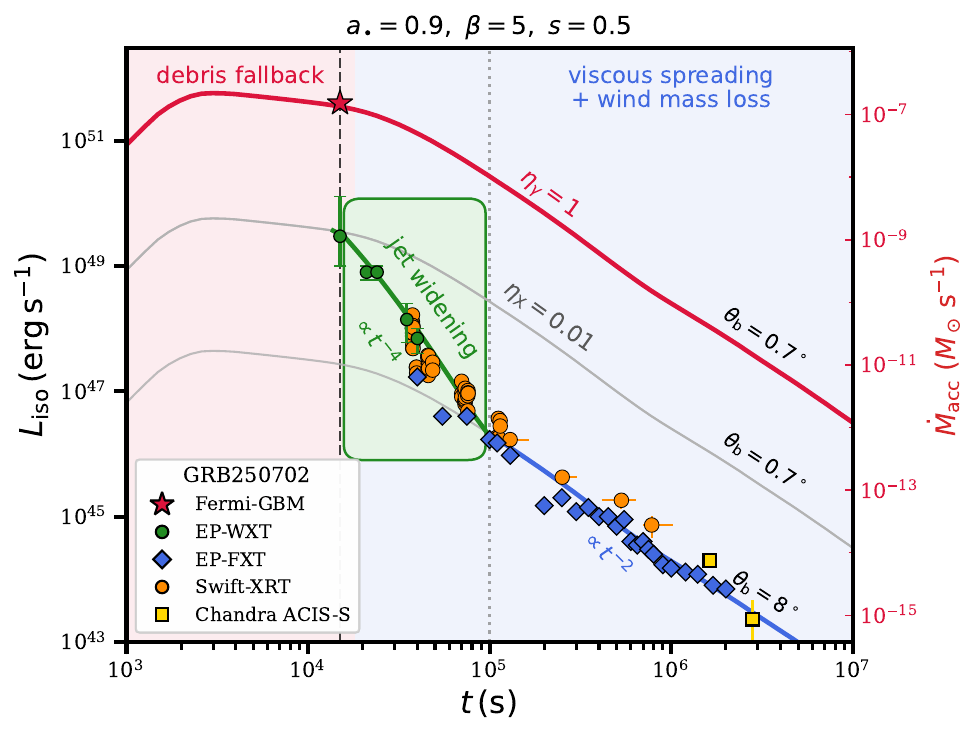}
\caption{
Our jetted micro-TDE engine model reproduces the light curve of
GRB\,250702 in three frequency bands: the tightly beamed relativistic jet
($\theta_{\rm b}=0.7^\circ$) powers the prompt $\gamma$-rays;
wind-modified accretion produces the late-time week-long EP-FXT tail
($L_{\rm iso}\propto t^{-2}$); and post-breakout jet widening
($\theta_{\rm b}\propto t$) results in the sharp EP-WXT X-ray decay at
the intermediate times (by steepening the dependence to
$L_{\rm iso}\propto t^{-4}$). The X-ray data (EP-WXT, green circles;
EP-FXT, blue diamonds) are obtained from \cite{EylesFerris2026}; the red
star is the isotropic equivalent $\gamma$-ray luminosity of GRB\,250702
from Fermi-GBM at $z=1.036$ \citep{Neights2026}. The \textit{Swift}-XRT X-ray points are obtained from the Living \textit{Swift}-XRT Point Source catalogue \citep[][]{Evans2023}, and the \textit{Chandra} points are
adopted from \citet{EylesFerris2026}.
Time is measured in the source rest frame from MJD\,60858.2 (2025 July 2, 04:48 UTC), the reference epoch of \citet{EylesFerris2026}, which we identify with the complete disruption of the star. 
The black dashed vertical line marks the Fermi-GBM trigger (GRB\,250702D), coincident with the WXT flux peak, at $t\approx1.5\times10^{4}$\,s ($\approx4$ rest-frame hours after the
disruption); the fallback ($\sim10^{3}$\,s) and viscous ($t_{\rm v}\approx4$\,hr) rise times set this delay, so the prompt emission coincides with the accretion peak. 
The curves show the isotropic luminosity in each band, differing in the jet beaming angle  and radiative efficiency.
All curves share the same engine ($s=0.5$, $\eta_{\rm jet}=1$, $a_\bh=0.9$). The red curve shows the accretion rate, $\dot{M}_{\rm acc}$,
along the right y-axis. The $\gamma$-ray curve (red) uses the narrow
prompt jet, $\theta_{\rm b}=0.7^\circ$ and $\eta_\gamma=1$; the WXT curve
(green) follows the widening jet, $\theta_{\rm b}=0.7^\circ\!-\!8^\circ$
at $\eta_{\rm X}=0.01$; and the FXT curve (blue) uses the jet with the
saturated opening angle, $\theta_{\rm b}=8^\circ$, at
$\eta_{\rm X}=0.01$. Solid gray lines show constant beaming angle
reference solutions ($\theta_{\rm b}=0.7^\circ$ and $8^\circ$ at $\eta_{\rm X}=0.01$). The gray dotted line marks the end of jet widening phase. During the WXT phase the widening jet
($\theta_{\rm b}\propto t$) combined with the declining engine power
($L_{\rm jet}\propto t^{-2}$) gives
$L_{\rm iso}\propto L_{\rm jet}/\theta_{\rm b}^2\propto t^{-4}$,
reproducing the steep WXT decay ($\alpha\approx3.8$); once
$\theta_{\rm b}$ saturates, the beaming correction is constant and the
light curve tracks the wind-modified accretion rate ($s=0.5$), decaying as $t^{-2}$ (FXT phase).}
    \label{fig:Liso}
\end{figure*}

We adopt a full-disruption micro-TDE model in which the observed delay ($\sim 1$\,day) between the initial soft X-ray emission and the onset of the $\gamma$-ray burst reflects the timescale required for the stellar debris stream to circularize and form a disk with an evacuated polar funnel (along the disk rotation axis, Figure~\ref{fig:snapshots}), and for a relativistic jet to drill its way through the funnel and escape. Our single-disruption framework produces the multi-wavelength light curve without requiring specific emission mechanisms or repeated stripping (partial disruption) events, as detailed below.

\subsection{Pre-disk-formation phase: the X-ray precursor}

Our model accommodates the early WXT X-ray precursor as powered by the pre-viscous-accretion phase, during which the stellar debris stream has not yet circularized and settled into a disk. During this phase, a small fraction of the bound material can be accreted onto the BH (through low-angular-momentum stream collision products) and launch an early jet without significant radiation-driven wind suppression (i.e., $s=0$). This is motivated by recent GRMHD simulations which show that relativistic jets can be launched even from accretion of gas with little or no angular momentum, provided the inflow carries magnetic flux \citep{Lalakos2024}: the magnetic field threading the BH extracts its rotational energy via the BZ mechanism, and the BH's frame-dragging spins up the surrounding gas, which then forms a rotationally-supported flow. 
Magnetic flux advected by the early stream-fed accretion can therefore launch the early jet powering the precursor, with the disk building up subsequently.
The observed peak precursor luminosity, $L_{\rm X,iso}\sim10^{46}\,{\rm erg\,s^{-1}}$ \citep{Li2026}, is five orders of magnitude below the prompt $\gamma$-ray luminosity \citep{Neights2026}, which we attribute to a highly magnetized jet in the MAD state. This luminosity contrast therefore points to a correspondingly lower fallback rate, $\dot M_{\rm fb}\ll10^{-5}\,M_\odot\,{\rm s^{-1}}$ during the precursor, consistent with only a small fraction ($\lesssim1\%$) of the bound debris falling directly onto the BH and lower radiation efficiency $\eta_{\rm X}\lesssim 0.01 $.

\subsection{Post-disk formation phase: Wind-modified disk accretion}
\label{sec:post-disk}

Figure~\ref{fig:Piso} shows the isotropic-equivalent jet power, $P_{\rm iso}\equiv L_{\rm iso}/\eta_{\rm rad}$, as a function of the intrinsic jet power, $L_{\rm jet}$, and beaming angle, $\theta_{\rm b}=1/\gamma_{\rm j}$, where $\gamma_{\rm j}$ is the jet Lorentz factor. Here, we identify the engine configurations that reproduce the peak $\gamma$-ray isotropic equivalent luminosity of GRB\,250702 ($L_{\gamma,\rm iso}\sim10^{51.6}$~erg~s$^{-1}$) for a range of $\gamma$-ray efficiency, $\eta_\gamma$ (solid and dashed red lines). The allowed parameter space is the wedge with $\Lambda>2$ and $\eta_\gamma\le1$: the jet must be kink-stable ($\Lambda>2$, cyan line) and powerful enough to produce the observed luminosity ($\eta_\gamma \le 1$, solid red line). The stability and energetics boundaries meet near our minimum power fiducial central engine, marked with the magenta star: it features high-power and strongly beamed jets, 
$L_{\rm jet}\approx3\times10^{47}$~erg~s$^{-1}$ and $\theta_{\rm b}=0.7^\circ$, respectively. Our model achieves this solution at the peak BH accretion rate,
$\dot M_{\rm acc}\sim10^{-7}\,M_\odot\,{\rm s}^{-1}$, for a rapidly spinning
BH, $a_\bh=0.9$, and fiducial disk wind mass loss slope, $s=0.5$.
While these fiducial parameters lie on the $\eta_\gamma = 1$ boundary, relaxing the wind mass loss to lower $s$ raises the accretion rate and opens up the radiatively-allowed parameter space ($\eta_\gamma < 1$, unshaded in Figure~\ref{fig:sa_plane}): a modestly weaker wind ($s \approx 0.3$) lowers the efficiency required to reproduce GRB\,250702B to $\eta_\gamma \approx 0.2$ at $a_\bullet = 0.9$, and lower spins become viable with $\eta_\gamma < 1$ for $s \lesssim 0.35$. Our model can thus reproduce GRB~250702B energetics over a broad range of spin and radiative efficiency for weaker winds ($s<0.5$).

\subsubsection{The EP-WXT phase: Jet widening}
\label{sec:widening}

Having established the wind-modified viscous accretion model in Section~\ref{sec:disk_model}, we now compare it to the light curve of GRB\,250702. Figure~\ref{fig:Liso} shows the $\gamma$-ray and X-ray isotropic-equivalent luminosity light curves, computed by numerically solving the coupled disk evolution equations (Equations~\ref{eq:Mdotd}--\ref{eq:Rdot}), with the fallback rate, $\dot{M}_{\rm fb}(t)$, taken directly from our \textsc{arepo} simulations. We adopt the parameters $a_\bh = 0.9$ (yielding jet efficiency, $\eta_{\rm jet}\approx 1$ for $\varphi_\bh = 50$), radiative efficiency $\eta_{\rm X} = 0.01$ and $\eta_{\gamma} = 1$, and wind exponent $s = 0.5$, accretion radius $r_{\rm acc}=10\,r_{\rm g}$ and initial beaming angle $\theta_{\rm b} = 0.7^\circ$ which together set the peak isotropic X-ray and $\gamma$-ray luminosity to $L_{\rm X,iso}\sim 10^{49}$\,erg\,s$^{-1}$ and  $L_{\gamma, \rm iso}\sim 10^{51}$\,erg\,s$^{-1}$, respectively \citep{Li2026,Neights2026}.

The horizontal axis shows the rest-frame time since the pericenter passage leading up to the full disruption of the star. We set $t=0$ at MJD~60858.2 (2025 July 2, 04:48 UTC), the reference epoch of \citet{EylesFerris2026}. Observationally, this epoch lies $\approx 1.9$\,hr in the observer frame ($\approx 0.9$\,hr in the rest frame) after the start of the first EP-WXT pointing in which the source is individually detected at 02:53:45 UTC \citep[Table S1;][]{Li2026}, and $\approx 4.1$\,hr (rest frame) before the first Fermi-GBM trigger (GRB~250702D); the subsequent triggers, GRB~250702B and E, occur at $t \approx 4.5$ and $5.7$\,hr, respectively. 
At early times, while the disk mass $M_{\rm d}$ is still small, $\dot M_{\rm fb} > M_{\rm d}/t_{\rm v}$ and the disk grows: the evolution is fallback-limited (shaded red region). Once most of the bound debris has returned to pericenter, the fallback supply drops below the viscous drain, the disk mass peaks, and the evolution becomes dominated by viscous accretion and wind mass loss (shaded blue region). In our fiducial model this transition occurs at $t \simeq 1.8\times10^4$\,s ($\simeq 5$\,hr; red-to-blue shading in Figure~5). The prompt $\gamma$-ray emission at $T_0 \simeq 4$\,hr thus coincides with the end of the fallback-limited phase, when the accretion rate, and hence the jet power, is near its maximum.
Stacking earlier WXT
observations, however, showed that the X-ray emission was already present on 2025
July 1 at 01:40:26 UTC \citep{Li2026}, $\approx 17$\,hr (rest frame) before the
first GBM trigger. If no mass is stripped during earlier pericenter passages, the
X-ray precursor lead time before the first GRB is set by the fallback and viscous accretion times
(Eq.~\ref{eq:tv}), which depend on the product of the viscosity
parameter and the square of the disk aspect ratio, $\alpha_{\rm v}(H/R)^2$.

In analogy with the fast- and slowed-accretion
scenarios of \citet{Li2026}, we distinguish two regimes governed by the viscous accretion
time. In the \textit{fast-accretion} case we adopted in Fig.~\ref{fig:Liso}
($\alpha_{\rm v}=0.1$, $H/R=0.5$), the viscous time is short, $t_{\rm v}\approx
4$\,hr, so the prompt $\gamma$-ray emission lags the onset of the rising X-ray
emission by $\approx 4$\,hr and is co-eval with the first individually-detected
WXT pointing post-$\gamma$-ray trigger; this case reproduces the observed multi-band light curve, including the 
late-time $t^{-2}$ decline slope, which is set by the wind parameter, $s=0.5$
(Eq.~\ref{eq:scaling3}). In the \textit{slow-accretion} case, less efficient
angular-momentum transport ($\alpha_{\rm v}\approx 0.025$) increases the viscous
time to $t_{\rm v}\approx 17$\,hr, matching the observed $\approx 17$\,hr (rest frame)
precursor-to-prompt delay. (Alternatively, the X-ray precursor may arise from
a partial stripping event during the star's previous pericenter passage before the subsequent complete
disruption triggers the prompt $\gamma$-ray emission \citep[e.g.,][]{EylesFerris2026}.)
In the time convention of Figure~\ref{fig:Liso}, where $t=0$ marks the complete
disruption and the prompt emission occurs at $t\approx 4$\,hr, the observed
$\approx 17$\,hr precursor-to-prompt delay places the precursor at $t\approx
-13$\,hr, corresponding to a partial stripping event on an earlier orbit $\approx 13$\,hr
before the final disruption.

The EP-WXT X-ray light curve (green circles) decays as $L_{\rm X,iso}\propto t^{-\alpha}$ with $\alpha \approx 3.8$, significantly steeper than the intrinsic engine decay, $L_{\rm jet}\propto t^{-2}$, predicted by our wind-modified accretion model for $s = 0.5$ (Eq.~\ref{eq:scaling3}). 
We propose that this steepening reflects a time-dependent jet beaming geometry. In our model, the jet beaming angle widens linearly with time, $\theta_{\rm b}\propto t$, starting from the time of breakout and continuing until $\sim 10^{5}$~s: as a result, $\theta_{\rm b}$ grows from $\theta_{\rm b,i} = 0.7^\circ$ to $\theta_{\rm b,f}= 8^\circ$. Because $\theta_{\rm b}\propto t$, the beaming fraction scales as $\theta_{\rm b}^{2}\propto t^{2}$ and---when combined with the intrinsic engine decay, $L_{\rm jet}\propto t^{-2}$---leads to the steep decay, $L_{\rm X,iso}\propto L_{\rm jet}/\theta_{\rm b}^{2} \propto t^{-4}$, observed
during the EP-WXT phase.

Such widening can arise from several physical mechanisms, including the reconnection-driven dissipation of the toroidal  magnetic field, which weakens the magnetic hoop stress that helps to collimate the jets \citep{Bromberg2016, Ripperda2022}, and the lateral expansion at the comoving sound speed once the cocoon disperses upon breakout~\citep{Bromberg2011, MizutaIoka2013, Harrison2018, 2010NewA...15..749T,2010MNRAS.407...17K}.  We note that the beaming evolution in the latter case is an early-time transient signature of the breakout rather than a property of a steady
jet. At any given time the observed emission originates from the part of the jet
near its photosphere rather than from the jet as a whole; the beaming correction
therefore evolves as the jet transitions from a cylindrical jet confined inside the envelope to a freely expanding one outside. Just before the jet exits the confining envelope, the jet beaming angle is its relativistic beaming angle, $\theta_{\rm b,i} = 1/\gamma_{\rm j}$. After the head of the jet breaks out of the envelope (and ceases to produce the backflows, which collimated the jet into a cylinder), the parts of the jet still inside the envelope start to expand laterally until they fill the low-density funnel of half-opening angle $\approx 15^\circ$ (Fig.~\ref{fig:snapshots}). Our best-fit final beaming angle, $\theta_{\rm b,f} = 8^\circ$ (Fig.~\ref{fig:Liso}), corresponds to the jet widening about halfway to filling the funnel. Several effects can prevent the jet from opening all the way to the funnel wall, such as the residual cocoon occupying the outer part of the funnel, so we regard $\theta_{\rm b,f}= 8^\circ$ as a natural upper limit to the jet opening angle set by the confining envelope geometry.  

A more detailed physical treatment is beyond the scope of this work; here we demonstrate that the observed $L_{\rm X,iso}\propto t^{-4}$ decline is reproduced by linear opening-angle growth in time, combined with the intrinsic $L_{\rm jet}\propto t^{-2}$ engine decay, which naturally comes out of our wind-modified accretion model.

\subsubsection{The EP-FXT phase: long-lived X-ray emission}
After the beaming angle saturates at $\theta_{\rm b}\sim 8^\circ$, the beaming fraction becomes constant, and the observed light curve tracks the intrinsic $L_{\rm jet}\propto t^{-2}$ decay from wind-modified viscous accretion, in agreement with the EP-FXT data (blue diamonds) in Figure~\ref{fig:Liso}. The FXT emission remains detectable for $\sim$\,weeks, far exceeding the viscous timescale evaluated at the initial circularization radius, $r_{\rm c}\simeq 1R_{\odot}$, for the deep encounter ($\beta=5$): $t_{\rm v}\approx4\,{\rm hr}$ for $\alpha_{\rm v}=0.1$, $h=0.5$  (Equation~\ref{eq:tv}).
Such a long emission tail is a natural consequence of the wind-modified viscous accretion: at $t\gg t_{\rm v}$, the disk enters the viscous spreading phase in which the accretion rate onto the BH decays as $\dot{M}_{\rm acc}\propto t^{-2}$ and the disk radius grows as $r_{\rm d}\propto t^{2/3}$ (Equation~\ref{eq:scaling1}). For instance, by $t\sim1$--$3$ weeks post-disruption, the disk spreads by a factor, $r_{\rm d}/r_{\rm c}\sim(t/t_{\rm v})^{2/3}\sim10$--$25$, which prolongs the viscous time at the disk outer edge accordingly, $t_{\rm v}(r_{\rm d})=t_{\rm v}(r_{\rm c})\,(r_{\rm d}/r_{\rm c})^{3/2}=1$--$3$~weeks. Thus, the wind-modified viscous accretion naturally explains the self-similar late-time power-law decline of the month-long (source frame) X-ray emission observed by EP-FXT/\emph{Swift}/\emph{Chandra}.

\section{Summary and Discussion}
\label{sec:conclusions}
\subsection{Summary}

We have studied jet formation and escape in micro-TDEs as a channel for powering ultra-long GRB\,s. Using 3D \textsc{arepo} hydrodynamic simulations of a $1\,M_\odot$ star disrupted by a
$10\,M_\odot$ BH, we showed that within $\lesssim1$\,day the debris settles into a disk with a low-density polar funnel along the angular momentum direction of the disk, with a steep $\rho\propto r^{-2}$ polar profile and half-opening angle $\theta_{\rm f}\approx15^\circ$. The two encounter geometries we considered ($\beta=1.4$ and $5$) both produce well-defined funnels, but the grazing encounter ($\beta=1.4$) develops a thinner, rotationally supported extended disk. Deeper encounter ($\beta=5$) instead produces more compact, thicker inner disk ($h\simeq 0.5$), yielding a higher accretion rate and correspondingly a higher jet power. The funnel geometry, combined with the steep $r^{-2}$ polar profile, provides ideal conditions for jet collimation and stable propagation and escape for both encounters. 

Coupling our hydrodynamic simulations to a semi-analytic wind-modified disk evolution model, we find that jetted micro-TDEs can match the energetics and timescales of ultra-long GRB\,s: their hours-to-days viscous accretion times set the engine duration, and a highly collimated ($\theta_{\rm j}\lesssim1^\circ$) powerful ($L_{\rm jet}\gtrsim 10^{47}{\rm erg\,s^{-1}}$) jet from a rapidly spinning BH produces the observed $\gamma$-ray flux. These results establish jetted micro-TDEs as a physically motivated ULGRB\, engine. Applying this framework to GRB\,250702B, we account for all three of its observed phases (the soft X-ray precursor, the several-hour prompt $\gamma$-ray emission, and the weeks-long X-ray decay) as a natural progression in the aftermath of a full disruption of a Sun-like companion star. We propose a stream-fed origin for the X-ray precursor and quantitatively reproduce the X-ray light curve without invoking external-shock afterglow physics. Our main results are as follows.

 We propose the following three emission phases: (i) the X-ray precursor traces pre-disk stream-fed accretion, with only a small fraction ($\lesssim1\%$) of the debris reaching the BH before disk formation; (ii) once the disk forms and clears out the polar direction, a narrowly collimated jet with a beaming angle $\theta_{\rm b} \lesssim 1^\circ$ launched by a spinning BH in the MAD state propagates through the funnel and powers the prompt burst, energetic enough to match the observed $L_{\gamma,\rm iso}\sim10^{51}\,{\rm erg\,s^{-1}}$; and (iii) the subsequent X-ray emission arises from wind-modified viscous drainage of the disk ($s=0.5$), with the engine power declining as $L_{\rm jet}\propto t^{-2}$ while gradual jet widening from $\theta_{\rm b}=0.7^\circ$ to $8^\circ$ produces the steep $L_{\rm X,iso}\propto t^{-4}$ decline observed by EP-WXT. At later times, the beaming angle saturates at $ 8^\circ$ and the intrinsic $t^{-2}$ decay is recovered, matching the temporal slope of the long-lived EP-FXT tail, whose weeks-long duration follows from the viscous disk spreading ($r_{\rm d}\propto t^{2/3}$).

The kink-instability analysis further establishes that the inferred micro-TDE jet power for GRB\,250702B sits at or above the jet stability threshold, with the stability criterion ($\Lambda > 2$) satisfied at $L_{\rm jet}\gtrsim 10^{47}\,{\rm erg\,s^{-1}}$ for $\theta_{\rm b}\lesssim 1^\circ$ in the deep encounter case and at much lower jet powers for the wider jet opening angles relevant to the FXT phase. The multi-phase evolution therefore proceeds entirely within the stable regime of the $\Lambda$--$L_{\rm jet}$ plane, confirming that the polar funnel geometry established during the disruption is favorable for launching stable, well-collimated relativistic jets throughout the emission window. Notably, the $r^{-2}$ polar profile sits precisely at the critical slope $\alpha=2$ that separates jet collimation from de-collimation \citep{Bromberg2016}. Because $\alpha=2$ implies $\rho_{\rm a}z_{\rm h}^2=\,$constant, $\Lambda$ is then independent of jet-head distance, so stability holds at all radii without requiring breakout at a particular height.

\subsection{Discussion}

Several progenitor scenarios have been proposed for GRB\,250702B, ranging from a single engine to distinct processes driving each phase: tidal disruption of a white dwarf by an intermediate-mass BH \citep{Li2026, EylesFerris2026, Sato2026}, an external-shock afterglow origin for the late X-rays \citep{OConnor2025}, self-regulated collapse of a supergiant star \citep{Zhang2026}, and micro- or milli-TDEs of main-sequence stars \citep{Beniamini2025, Granot2025}. We compare our jetted micro-TDE model to each of these previous studies below, focusing on the physical mechanisms governing the different phases of GRB\,250702B.

\paragraph{Fallback versus wind-driven accretion} 

At the highly super-Eddington rates expected in micro-TDEs, the geometrically thick, advection-dominated flow ejects most of the inflowing mass as winds before it reaches the BH \citep[e.g.,][]{Narayan1994, Blandford1999}, so that only a small fraction ($\lesssim1\%$) of the bound debris is ultimately accreted. A key feature of our model relative to previous TDE interpretations of GRB\,250702B \citep[e.g.,][]{Beniamini2025,Granot2025} is that the accretion rate is evolved self-consistently with a wind-loss prescription: $\dot{M}_{\rm acc}\propto (r_{\rm acc}/r_{\rm d})^{s}$ with $s=0.5$. The resulting suppressed accretion rate, $\dot{M}_{\rm acc}\sim  10^{-7}\,M_\odot\, {\rm s}^{-1}$, sets the scale of the peak luminosity: considering disk accretion alone results in isotropic X-ray luminosities of up to $\sim10^{46}\,{\rm erg\,s^{-1}}$ for $s=0.5$ \citep[e.g.,][]{Kremer2022}, several orders of magnitude below the isotropic $\gamma$-ray luminosity of GRB\,250702B. Reproducing the prompt isotropic equivalent $\gamma$-ray luminosity ($L_{\gamma,\rm iso}\sim 10^{51}\,{\rm erg\,s^{-1}}$) at these suppressed accretion rates therefore requires beamed jet emission rather than quasi-isotropic disk radiation: a narrowly collimated jet ($\theta_{\rm b}\lesssim1^\circ$, beaming factor $f_{\rm b}^{-1}\sim10^{3}$--$10^{4}$) launched by a spinning BH via the Blandford--Znajek mechanism.

\paragraph{Engine duration and recurrence.} In our model, the $\sim$hours-long prompt $\gamma$-ray duration is set by the viscous timescale of an extended ($\sim R_\odot$) micro-TDE disk (Eq.~\ref{eq:tv}), whereas collapsar scenarios invoke extended envelope fallback \citep{Zhang2026} and the intermediate-mass BH scenarios tie the timescale to the larger BH mass through the fallback time of a disrupted white dwarf \citep{Li2026, EylesFerris2026} or main-sequence star \citep{Granot2025}. A further distinction of our model is that we incorporate viscous spreading of the disk, which regulates the late-time emission by extending the engine lifetime: the growth of the disk ($r_{\rm d}\propto t^{2/3}$), combined with the wind-modified accretion ($\dot{M}_{\rm acc}\propto t^{-2}$), extends the local viscous time at the disk edge to weeks. 
The origin of the quasi-regular recurrence of the prompt episodes remains uncertain: \citet{An2025} and \citet{Sato2026} attribute it to Lense--Thirring precession of a compact, misaligned inner torus at
$r\sim10^2$--$10^3\,r_{\rm g}$, and \citet{Zhang2026} to the collapse of successive stellar layers. In our model the debris disk is radially extended ($\sim10^5\,r_{\rm g}$), and its large angular momentum content makes global Lense--Thirring precession inefficient, with precession periods far exceeding the engine lifetime; we therefore do not expect global disk precession under spin--disk misalignment. Instead, the recurrence may reflect ongoing debris fallback perturbing the disk and modulating the jet. We regard this as a qualitative possibility; confirming it requires coupled GRMHD modeling of the jet--disk coupling, which we defer to future work.

\paragraph{Late-time thermal component.} At late times, EP250702a, produces a thermal component, $kT_{\rm obs} \approx 170$~eV, at an isotropic equivalent luminosity, $L_{\rm obs,iso} \sim 10^{44}\,{\rm erg\,s^{-1}}$~\citep{Li2026, Chen2026}. In our model this emission originates from the same jet-driven outflow, with its evolution from hard $\gamma$-rays to the soft thermal X-ray emission following from the evolving jet geometry. While confined within the debris envelope, the jet is nearly cylindrical with a very small opening angle
($\gamma_{\rm j}\theta_{\rm j} \approx 0.015 \ll 1$); such a jet is optically thick, so its photosphere lies within the relativistic flow and the escaping jet-frame X-ray emission is Doppler-boosted by the relativistic jet motion (Lorentz factor $\gamma_{\rm j}\sim100$) to a few hundred keV, resulting in hard $\gamma$-ray emission. After breakout, the jet widens and becomes transparent, so the emission instead emerges from the irradiated, the non-relativistic jet-drilled funnel walls, which produce a softer X-ray component. We can estimate the funnel-wall effective temperature,
\begin{equation}
T_{\rm eff} =
\left[\frac{L_{\rm jet,iso} f_{\rm b}/2}
{\sigma_{\rm SB} \pi (z\theta_{\rm j})^2}\right]^{1/4}
\approx  \left[\frac{L_{\rm jet,iso}}{4\pi\sigma_{\rm SB}z^2}\right]^{1/4},
\label{eq:Teff}
\end{equation}
where $\sigma_{\rm SB}$ is the Stefan--Boltzmann constant, $f_{\rm b} \approx \theta_{\rm j}^2/2$ is the beaming fraction, the factor of $1/2$ converts the twin-jet luminosity to a single jet, and $z$ is the distance to the emission region. Although the envelope remains scattering-thick out to distances of hundreds of $R_\odot$, the relevant surface is the effective photosphere of the wall, which lies much deeper, near the stellar-scale inner disk in our simulations. Figure~\ref{fig:Liso} shows that our model gives an isotropic equivalent jet luminosity, $L_{\rm jet,iso} \sim 10^{44}\,{\rm erg\,s^{-1}}$, at the time of the thermal X-ray detection of $\sim 10$ days.
For this luminosity, eq.~\eqref{eq:Teff} gives $T_{\rm eff} \sim 60$--$200$\,eV,
bracketing the observed value ($kT_{\rm obs} \approx 170$\,eV), for the emission distance ranging from
$z \sim 10\,R_\odot$ (the breakout scale) to $z \sim 1\,R_\odot$ (the
inner-disk/circularization scale).

\textit{Jet geometry and the radio counterpart.} ---The one-year radio
monitoring of GRB\,250702B by \citet{Goodwin2026}, and in particular the absence of
a clear jet break, allows two distinct jet geometries that the radio data alone
cannot distinguish: a narrow, highly relativistic jet
($\theta_{\rm j}\lesssim 2^\circ$, with a jet break already past), or a
wide-angle, mildly relativistic jet ($\theta_{\rm j}\gtrsim 15^\circ$,
$\Gamma\lesssim 10$) carrying a large kinetic energy
($E_{\rm K}\sim 10^{50}$\,erg) with no jet break out to $\sim355$~days. Our
micro-TDE model prefers the former. We find from our \textsc{arepo} simulations
that the debris envelope forms an evacuated polar funnel with a half-opening
angle of $\approx 15^\circ$, which sets the maximum jet opening angle and
therefore disfavors the mildly relativistic, wide-jet scenario. The energy
budget weighs against it as well: even for no wind mass loss
($s=0$; the top horizontal axis of Fig.~\ref{fig:Piso}), the maximum jet power
$L_{\rm jet}\sim \dot{M}_{\rm acc}c^2 \approx 3\times10^{49}$\,erg\,s$^{-1}$ falls
short of the $\sim 10^{50}$\,erg\,s$^{-1}$ of
intrinsic power needed to produce the observed $L_{\gamma,\rm iso}$ at
$\theta_{\rm j}\approx 15^\circ$. Although a larger viscosity parameter, $\alpha_{\rm v}\gtrsim 0.1$, increases the
peak accretion rate and, hence, $L_{\rm jet}$, it also shortens the viscous time
($t_{\rm v}\propto\alpha_{\rm v}^{-1}$; Eq.~\ref{eq:tv}) to below $\approx 4$~hr---shorter than both the rise of the observed X-ray light curve to its peak and the duration of the prompt $\gamma$-ray emission---inconsistent with observations.

The same wind parameter that governs this
energy budget also sets the rate of the late-time decay: our fiducial $s=0.5$ reproduces the
observed FXT decline $L_{\rm X,iso}\propto t^{-2}$ through the mass loss that
steepens the accretion rate and thus jet power. The fit to the late-time light curve therefore supports a low-energy engine, which together with the funnel geometry points to a narrow, relativistic jet that naturally produces an early jet break. This breaks the degeneracy left open by the radio data and motivates early
multi-wavelength follow-up of future ULGRBs.

  Our model captures the jet interactions with the surrounding outflows over a wide range of scales. On the larger, resolved scales ($r \gtrsim 0.1\,R_\odot$), non-axisymmetric modes in our \textsc{arepo} simulations manifest themselves via an effective viscosity, analogous to the $\alpha$-parameter of the standard disk theory, and naturally produce envelope-powered outflows. These outflows carve out the funnel and determine the radial density profiles that we use as inputs to our jet-propagation model. We note that the \textsc{arepo} simulations include the radiation pressure through the equation of state.
  We neglect the magnetic effects on the large scales and defer them to future work: their main effect is to quantitatively increase the effective viscosity that drives accretion.
  On the smaller, unresolved scales
($r_{\rm H} \lesssim r \lesssim 0.1\,R_\odot$), our analytic disk model accounts for the disk wind in two ways: it reduces the mass inflow reaching the BH (parametrized by the power-law slope, $s$) and provides the pressure support that collimates and accelerates the jets to $\gamma_{\rm j} \sim 100$ with $\theta_{\rm j} \sim 1/\gamma_{\rm j}$ (until the jets run into the envelope and collimate into cylinders). The relativistic jet properties in our model thus originate from the collimation by the disk winds.

\textit{Failed jets and low-luminosity ULGRBs.} ---Because GRB\,250702B lies close to the kink-stability threshold, a larger population of events below it, powered by lower accretion rates or more slowly spinning black holes, should exist. An analogy with low-luminosity GRBs \citep[e.g.,][]{Liang2007} points to what such engines could produce. \citet{Bromberg2011b} showed that low-luminosity GRBs form a distinct population whose jets cannot break out of the progenitor star within their observed durations. The central engine does not operate long enough to drive the jet through the star, so the shock and cocoon breakout instead powers the emission. In the ULGRB case the reason a weak jet fails to break out is different. Because the engine lifetime is intrinsically long, the breakout-time requirement is not constraining; instead, a sub-threshold jet fails because it is too under-powered to remain stable against the current-driven kink instability and disrupts before escaping the envelope. Following this analogy, we expect a population of low-luminosity ULGRBs powered by the cocoon that bubbles out through the funnel and drives shocks into the surrounding debris.
These transients would be fainter and more weakly collimated than ULGRBs, and, being shock/cocoon powered, their emission might be shorter and/or more intermittent than the prompt duration of a successful ULGRB. Characterizing these events and identifying how to find them is an open question, and motivates future GRMHD simulations of weaker ULGRB jets.

Several aspects of our model warrant further investigation. Dedicated GRMHD
  simulations will determine under what conditions a MAD state develops, as well as the accretion rate and jet power for a given BH spin. Such simulations can also test the jet widening to which we attribute the $t^{-4}$ decay observed by the EP-WXT (Sec.~\ref{sec:widening}).
  These simulations are also required to clarify the earliest phase of jet breakout: our jet-stability analysis uses the polar density profiles at the epoch when the funnel is established, but at earlier times ongoing fallback and outflows leave the funnel asymmetric, which could temporarily choke the jet and imprint observable signatures such as jet wobbling and quasi-periodic variability before the funnel clears and the jet escapes. In GRB\,250702B, these early asymmetries provide a natural origin for the episodic prompt emission: the three quasi-regularly spaced $\gamma$-ray episodes could mark intermittent choking and re-emergence of the jet as it drills through a funnel still being cleared by ongoing fallback. 

Finally, while the large-scale jet propagation considered here is not expected to depend strongly on the orientation of the BH spin, the effects of spin--disk misalignment on the inner accretion flow and jet launching warrant further investigation. In particular, strongly tilted systems ($\gtrsim 60^\circ$) may be less likely to reach the MAD state, potentially affecting the accretion rate and jet power \citep[e.g.,][]{Chatterjee2025}. Future GRMHD simulations could therefore test how the degree of misalignment influences MAD formation and the resulting jet properties. Such simulations would also allow us to explore the dynamics of bent jets propagating through the asymmetric debris environment and whether disk--spin misalignment introduces additional time variability into the jet emission. These effects could provide further observational signatures of the micro-TDE scenario beyond those considered in this work.

\begin{acknowledgments}
 
 We thank the anonymous referee for their insightful suggestions, which have helped to improve the manuscript. We thank Bart Ripperda, Jamie Lombardi, Omer Brom\-berg, Yuri Levin and Wen-fai Fong for  helpful discussions. We acknowledge the use of Anthropic’s Claude (Opus 4.8) for feedback on improving the clarity and readability of the text; all scientific content and conclusions are the authors' own. This work made use of data supplied by the UK \textit{Swift} Science Data Centre.
    F.K.\ acknowledges support from CIERA Postdoctoral Fellowship. 
    This work used computing resources at the Quest high-performance computing facility provided by CIERA under NSF Grant PHY-2406802.
    Quest is jointly supported by Northwestern University's Office of the Provost, the Office for Research, and Northwestern University Information Technology. K.K. acknowledges support from the Hellman Fellowship at UC San Diego.
    D. T. is supported by Harvard University through the Institute for Theory and Computation Fellowship.  B.D.M.\ acknowledges support from NASA (80NSSC24K0934) and the NSF (AST-2406637).  The Flatiron Institute is supported by the Simons Foundation.
    AT acknowledges support by NASA
80NSSC26K0343,
80NSSC22K0031,
80NSSC22K0799,
80NSSC18K0565 
and 80NSSC21K1746
grants, and by the NSF 
AST-2009884,
AST-2107839,
AST-1815304,
AST-1911080,
AST-2206471,
AST-2407475,
AST-2510570
grants.

\end{acknowledgments}

\appendix

\section{Disk properties}
\label{sec:disk_appendix}

\begin{figure}
    \centering
    \includegraphics[width=0.4\linewidth]{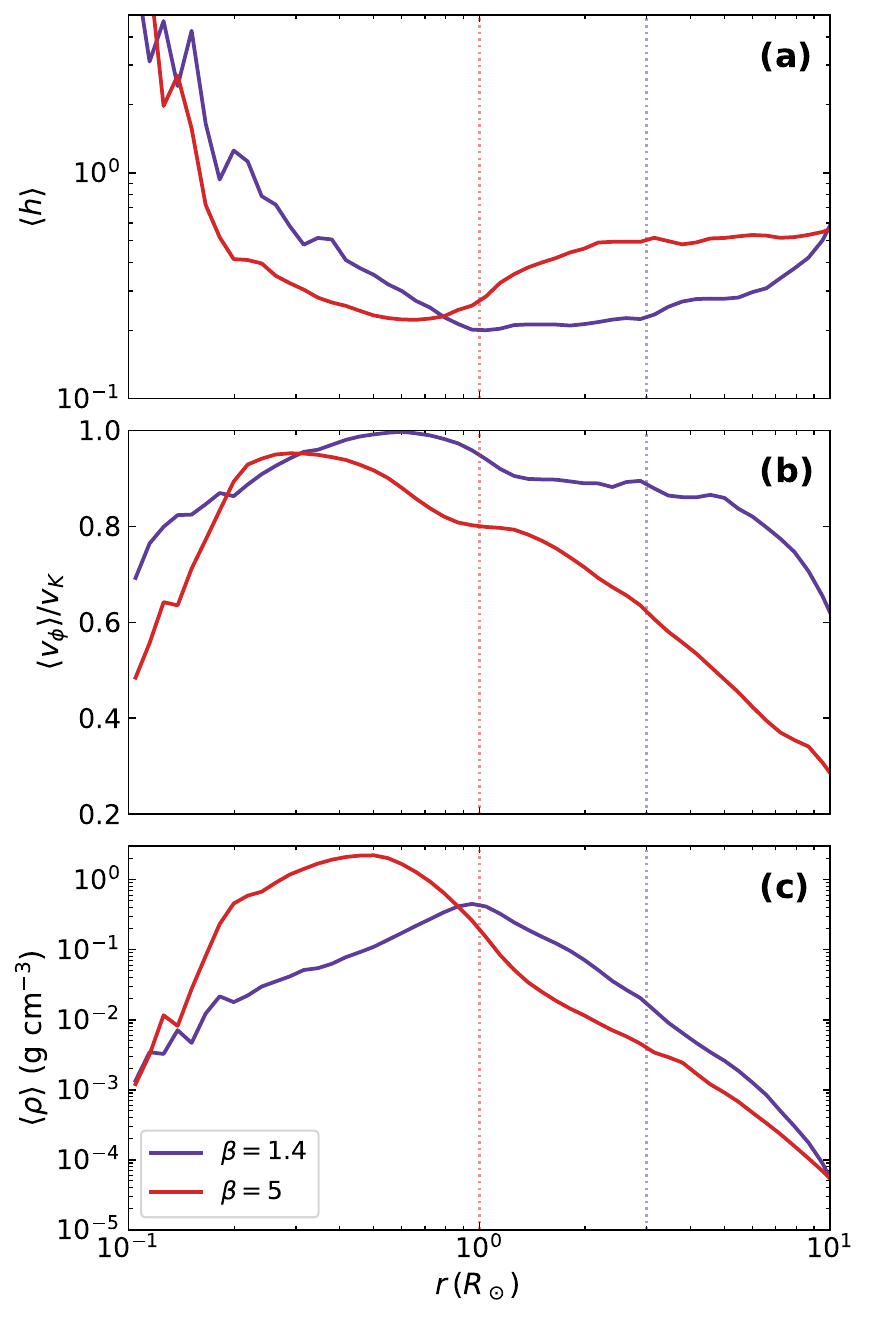}
    \caption{
    Radial profiles of the accretion disks formed by grazing ($\beta = 1.4$, purple) and deep ($\beta = 5$, red) encounters in our \textsc{arepo} simulations at $t = 4$\,hours post- disruption. \textbf{[panel (a)]:} disk aspect ratio $h$, showing that the two encounters settle into disks with plateau values $h \approx 0.5$ (deep) and $h \approx 0.2$ (grazing). \textbf{[panel (b)]:} ratio of the mass-weighted azimuthal velocity in the midplane ($|z|\leq h$) to the local Keplerian velocity, showing that the disk in the grazing case is rotationally supported out to a few $R_\odot$, while the deep encounter case becomes increasingly sub-Keplerian ($v_\phi/v_K \lesssim 0.8$) at $r \gtrsim 1\,R_\odot$. \textbf{[panel (c)]:} average midplane density (within $|z|\leq h$), with the deep encounter resulting in higher densities at small radii. Vertical dotted lines represent the circularization radius $r_{\rm c} \approx 2\,r_{\rm p}$, which serves as the initial disk radius for the viscous disk evolution model in Section~\ref{sec:tde}.
    }
    \label{fig:disk_profiles}
\end{figure}

Figure~\ref{fig:disk_profiles} shows the aspect ratio $h$, the ratio of the mass-weighted midplane azimuthal velocity to the local Keplerian velocity $v_\phi/v_K$, and the average midplane density as a function of radius, for both encounters at $t = 0.5$\,day. We define (a) the aspect ratio $h \equiv H/R$, where $H$ is the mass-weighted first-moment scale height,

\begin{equation}
H(R) = \frac{\sum_i m_i\,|z_i|}{\sum_i m_i},
\label{eq:h_def}
\end{equation}
with the sum taken over cells contained in a cylindrical shell of radius $R$ and width $\Delta \log R = 0.05$; (b) the ratio of the mass-weighted azimuthal velocity in the midplane region ($|z| < h$) to the local Keplerian velocity,
$\langle v_\phi\rangle/v_K$, where 
\begin{equation}
v_K = \sqrt{\frac{G[M_{\rm enc}(<R) + M_\bh]}{R}}
\label{eq:vk}
\end{equation}
uses the enclosed mass including the black
hole and the gas; and (c) the volume-weighted mean midplane density.

The bulk of the disk mass is concentrated around the circularization radius, $r_{\rm c }= 1$--$3R_\odot$, where the flow settles into a rotationally supported configuration with $v_\phi/v_K \approx 0.8$--$1$ and $h \approx 0.2$--$0.5$. The two encounters produce disks with different rotation profiles: the grazing case remains close to Keplerian rotation ($v_\phi/v_K \gtrsim 0.8$) out to $r \approx 3R_\odot$ with a thin disk ($h \approx 0.2$), whereas the deep case becomes progressively sub-Keplerian ($v_\phi/v_K \lesssim 0.8$) at $r \gtrsim 1R_\odot$ and vertically puffs up to $h \approx 0.5$. The deep encounter produces a centrally concentrated density distribution peaking at $\rho \approx 2$\,g\,cm$^{-3}$ near $r = 0.5R_\odot$, while the grazing encounter develops a broader, lower peak of $\rho \approx 0.2$\,g\,cm$^{-3}$ near $r = 1R_\odot$.
These structural differences reflect the depth of the pericenter passage: stronger tidal compression in the deep encounter drives  more efficient shock dissipation, concentrates mass at small radii, and reduces the rotational support in the outer flow. In contrast, the grazing encounter retains a larger fraction of its orbital energy and angular momentum, and circularizes more gradually at larger radii.

\begin{figure*}
    \centering
    \includegraphics[width=0.6\linewidth]{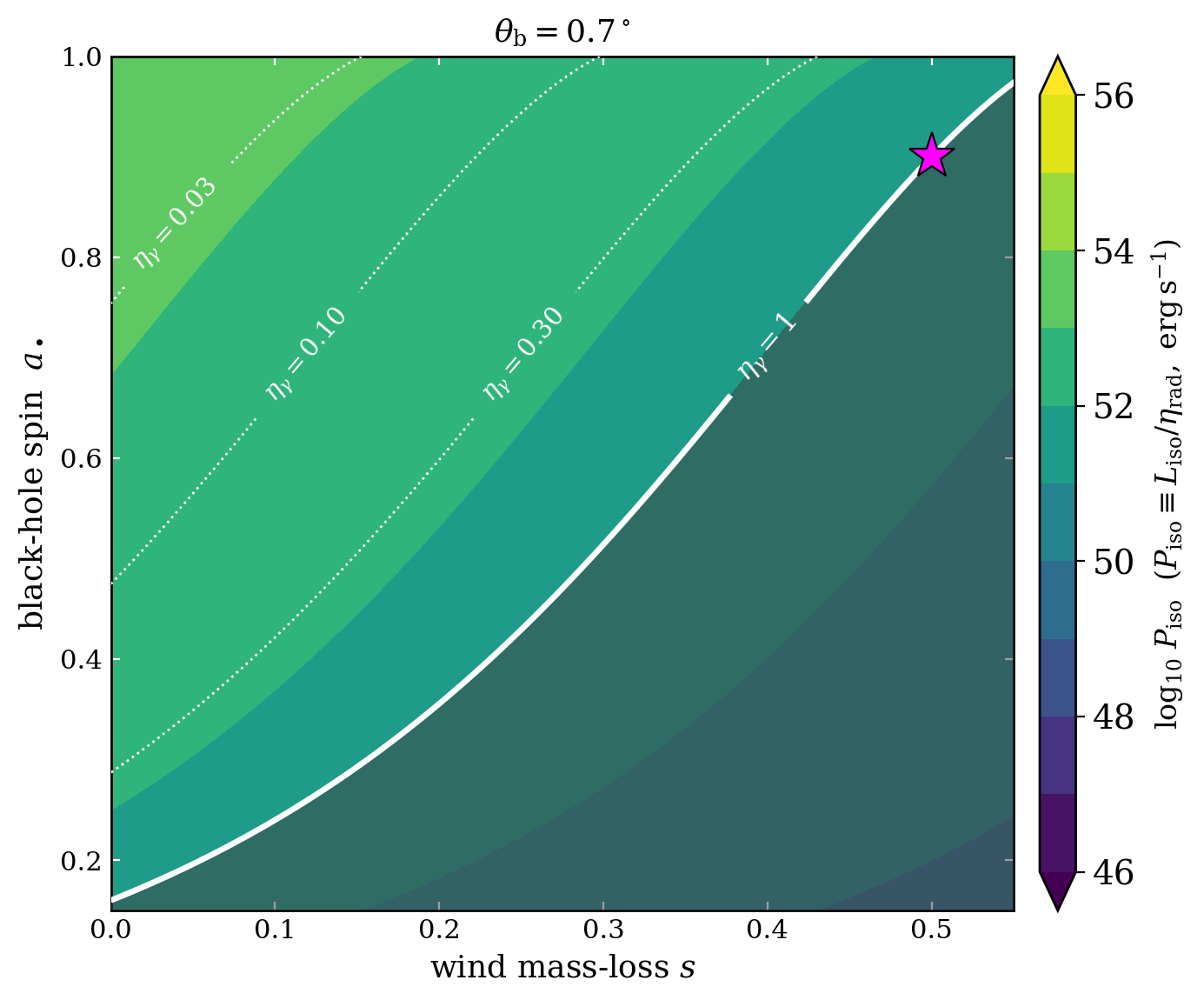}
\caption{Required $\gamma$-ray radiative efficiency $\eta_\gamma$ to reproduce GRB~250702B energetics across the wind-parameter--spin $(s, a_\bh)$ plane at fixed $\theta_{\rm b} = 0.7^\circ$ and $L_{\gamma,\rm iso} = 10^{51.6}\,{\rm erg\,s^{-1}}$. The color shows $P_{\rm iso}$; white contours mark $\eta_\gamma = 0.03$, $0.1$, $0.3$, and $1$. The magenta star is the fiducial parameter choice ($s = 0.5$, $a_\bh = 0.9$), which lies on
the $\eta_\gamma = 1$ boundary. Relaxing the wind mass loss to lower $s$ opens up a broad radiatively-allowed region ($\eta_\gamma < 1$), accommodating a wider range of BH spin $a_{\bh}<0.9$}
\label{fig:sa_plane}
\end{figure*}

\bibliographystyle{aasjournal}
\bibliography{mybib}
\end{document}